\documentclass[aps,pra,twocolumn,reprint,superscriptaddress]{revtex4-1}
\usepackage{amsfonts, amssymb, amsmath,graphicx}
\usepackage{subfigure}
\usepackage{multirow}
\usepackage{dcolumn}
\usepackage{bm}
\usepackage{color}

\begin{document}

\title{Klein tunneling in driven-dissipative photonic graphene}

\author{Tomoki Ozawa}
\affiliation{INO-CNR BEC Center and Dipartimento di Fisica, Universit\`a di Trento, I-38123 Povo, Italy}%
\author{Alberto Amo}
\affiliation{Centre de Nanosciences et de Nanotechnologies, CNRS, Univ. Paris-Sud, Universit\'e Paris-Saclay, C2N-Marcoussis, 91460 Marcoussis, France}%
\author{Jacqueline Bloch}
\affiliation{Centre de Nanosciences et de Nanotechnologies, CNRS, Univ. Paris-Sud, Universit\'e Paris-Saclay, C2N-Marcoussis, 91460 Marcoussis, France}%
\author{Iacopo Carusotto}
\affiliation{INO-CNR BEC Center and Dipartimento di Fisica, Universit\`a di Trento, I-38123 Povo, Italy}%

\date{\today}

\def\simge{\mathrel{%
         \rlap{\raise 0.511ex \hbox{$>$}}{\lower 0.511ex \hbox{$\sim$}}}}
\def\simle{\mathrel{
         \rlap{\raise 0.511ex \hbox{$<$}}{\lower 0.511ex \hbox{$\sim$}}}}
\newcommand{\feynslash}[1]{{#1\kern-.5em /}}
\newcommand{\iac}[1]{{\color{red} #1}}

\begin{abstract}
We theoretically investigate Klein tunneling processes in photonic artificial graphene. Klein tunneling is a phenomenon in which a particle with Dirac dispersion going through a potential step shows a characteristic angle- and energy-dependent transmission. We consider a generic photonic system consisting of a honeycomb-shaped array of sites with losses, illuminated by coherent monochromatic light. We show how the transmission and reflection coefficients can be obtained from the steady-state field profile of the driven-dissipative system.
Despite the presence of photonic losses, we recover the main scattering features predicted by the general theory of Klein tunneling. Signatures of negative refraction and the orientation dependence of the intervalley scattering are also highlighted. 
Our results will stimulate the experimental study of intricate transport phenomena using driven-dissipative photonic simulators.
\end{abstract}

\maketitle

\section{Introduction}

Lattices of photonic resonators have recently emerged as versatile simulators of condensed matter physics phenomena~\cite{Hartmann2016}. Thanks to the great flexibility and onsite control in their fabrication, photonic systems have opened a way to the investigation of phenomena otherwise difficult to access in conventional condensed matter systems.
In particular, pioneered by the study of the Bose-Einstein condensation~\cite{Kasprzak2006}, exciton polaritons have appeared as a very successful system in the research on quantum fluids of light~\cite{Carusotto2013}, demonstrating various manybody phenomena such as supefluidity~\cite{Amo2009}, Josephson oscillations~\cite{Abbarchi2013}, and the localization in flat bands in suitably designed lattice geometries~\cite{Baboux2016}. Other photonic platforms, such as microwave cavities~\cite{Wang2009}, silicon resonators~\cite{Hafezi2013b}, and arrays of coupled waveguides~\cite{Rechtsman2013b,Rechtsman2013a} made excellent progress in the observation of  topologically protected edge states.

Since its discovery, graphene, with its characteristic linear Dirac-like dispersion of electrons, has been the subject of intense study in solid-state physics~\cite{CastroNeto2009}. Along with these developments, there have been several studies on the simulation of graphene physics using photonics. Honeycomb lattice structures to simulate the physics of graphene have been realized in microwave cavities~\cite{Bittner:2010, Bellec:2013, Bellec:2014}, propagating waveguides~\cite{Rechtsman:2013PRL, Rechtsman2013a, Plotnik:2014}, and exciton-polariton microcavities~\cite{Kusudo:2013, Jacqmin2014, Milicevic:2015, Milicevic:2016}. All of these experimental realizations have been successfully modeled using tight-binding Hamiltonians whose band structure presents the characteristic massless Dirac cones responsible for a number of transport phenomena in graphene. However, an important difference between the solid-state (electronic) graphene and photonic graphene is that, apart from the carrier being fermionic or bosonic, photonic cavity systems are intrinsically dissipative and what is observed in experiments is typically the steady-state configuration resulting from the interplay of pump and losses~\footnote{This is not the case of light propagation experiments in waveguide arrays that directly simulate a time-evolution problem~\cite{Rechtsman:2013PRL, Rechtsman2013a, larre2015}. A typical limitation of these experiments is however the difficulty of getting an efficient energy-selectivity~\cite{straw}.}.

On the one hand, dissipation poses certain challenges upon simulating dynamical properties, such as transport phenomena, because photons can be lost during propagation. Alternative and more sophisticated experimental schemes are therefore required to address dynamical properties of graphene with photonic simulators.
On the other hand, the dissipative nature of photons can be an asset when comparing photonics to other kinds of simulators. For example, in photonic resonators, dissipation mostly takes place via the radiative escape of photons, which carry along with them complete information on the in-cavity photonic wave function. Furthermore, the driven-dissipative nature of these systems has opened a way to explore entirely new physics such as the dissipative phase transitions~\cite{Baumann2010, Klinder2015, Carusotto2005, Altman2015, Hartmann2008, Gerace2009, Carusotto2009, Tomadin2010, Fink2017, Fitzpatrick2017, Rodriguez2016}, the dissipative measurement of band topology~\cite{Ozawa2014, Bardyn2014}, and emergence of novel topologial states~\cite{Peano2016a, Peano2016b}.

In this paper, we investigate how driven-dissipative photonic systems can be used to simulate Klein tunneling, which is a characteristic transport phenomenon of graphene. In this effect, a particle with relativistic dispersion normally incident on a potential step perfectly transmits into the step independently of the step height. When the step is higher than the energy of the incident particle, the transmission into the step corresponds to a ``particle-like" state transforming into a ``hole-like" state. Klein originally discussed such a phenomenon in the context of relativistic quantum mechanics~\cite{Klein1929, Baym}, but an analogous phenomenon can also occur in graphene where electrons have a linear (relativistic) dispersion and obey a two-dimensional Dirac-like equation~\cite{Katsnelson2006, Cheianov2006, Pereira2006, Beenakker2008, Allain2011}. Klein tunneling in graphene has been experimentally observed through transport measurements where engineered $npn$ junctions provide potential barriers~\cite{Stander2009, Young2009, Lee:2015, Gutierrez:2016}. 
Klein tunneling of photons obeying the one-dimensional Dirac equation has also been theoretically analyzed ~\cite{Otterbach2009, Longhi2010, Esposito2011} and experimentally realized in coupled optical waveguides~\cite{Dreisow2012}.
However, the direct experimental realization of the two-dimensional Klein tunneling in a step configuration, instead of a barrier, is missing. In this photonic context, Klein tunneling has also been theoretically discussed for optical metamaterials~\cite{Guney2009}, propagating photonic waveguides~\cite{Bahat-Treidel2010, Bahat-Treidel2011}, and optical microcavities in the presence of spin-orbit coupling~\cite{Solnyshkov:2016}, but the lossy nature of photons was not taken into account. 

In this work, we show that the Klein tunneling can be directly observed also in a driven-dissipative model of photonic graphene based on a honeycomb-shaped array of coupled semiconductor microcavities. By taking advantage of the flexibility in designing the pump profile, transmission and reflection rates can be quantitatively evaluated from the steady-state profile of the light emitted by the lattice under a coherent monochromatic pump. The direct access to the real-space wave functions allows the observation of negative refraction. Our work demonstrates that the finite linewidth associated with losses in photonic devices does not significantly affect the phenomenon of Klein tunneling, but rather offers a useful means to experimentally simulate its microscopic details. While our discussion is focused to the specific case of polaritons in laterally patterned planar microcavities~\cite{Jacqmin2014, Milicevic:2015, Milicevic:2016}, it straightforwardly extends to other related systems such as microwave~\cite{Wang2009, Bellec:2013} and superconductor resonators~\cite{Houck2012}.
This work is the first step towards the study of Klein tunneling, negative refraction, and Veselago lensing in the presence of interactions, directly accessible in exciton-polariton lattices~\cite{Tanese2013}.

The structure of the article is the following. In Sec.~\ref{sec:basics} we briefly review the basics of Klein tunneling in systems without losses and in Sec.~\ref{photgra} we summarize the main consequences of the honeycomb geometry on the coherent pumping and on the intracavity field imaging from the emitted light. In Sec.~\ref{sec:schemes} we propose experimentally viable schemes to observe Klein tunneling effects and we make use of numerical simulations of a driven-dissipative tight-binding model to characterize the efficiency of our proposal in the ideal case of a large surface sample with small losses. In Sec.~\ref{sec:smaller} we then discuss how the main qualitative features of Klein tunneling survive when more realistic samples are considered with smaller spatial size and larger losses. Conclusions are finally drawn in Sec.~\ref{sec:conclu}.

\section{Klein tunneling}
\label{sec:basics}

\begin{figure}[htbp]
\begin{center}
\includegraphics[width=8.0cm]{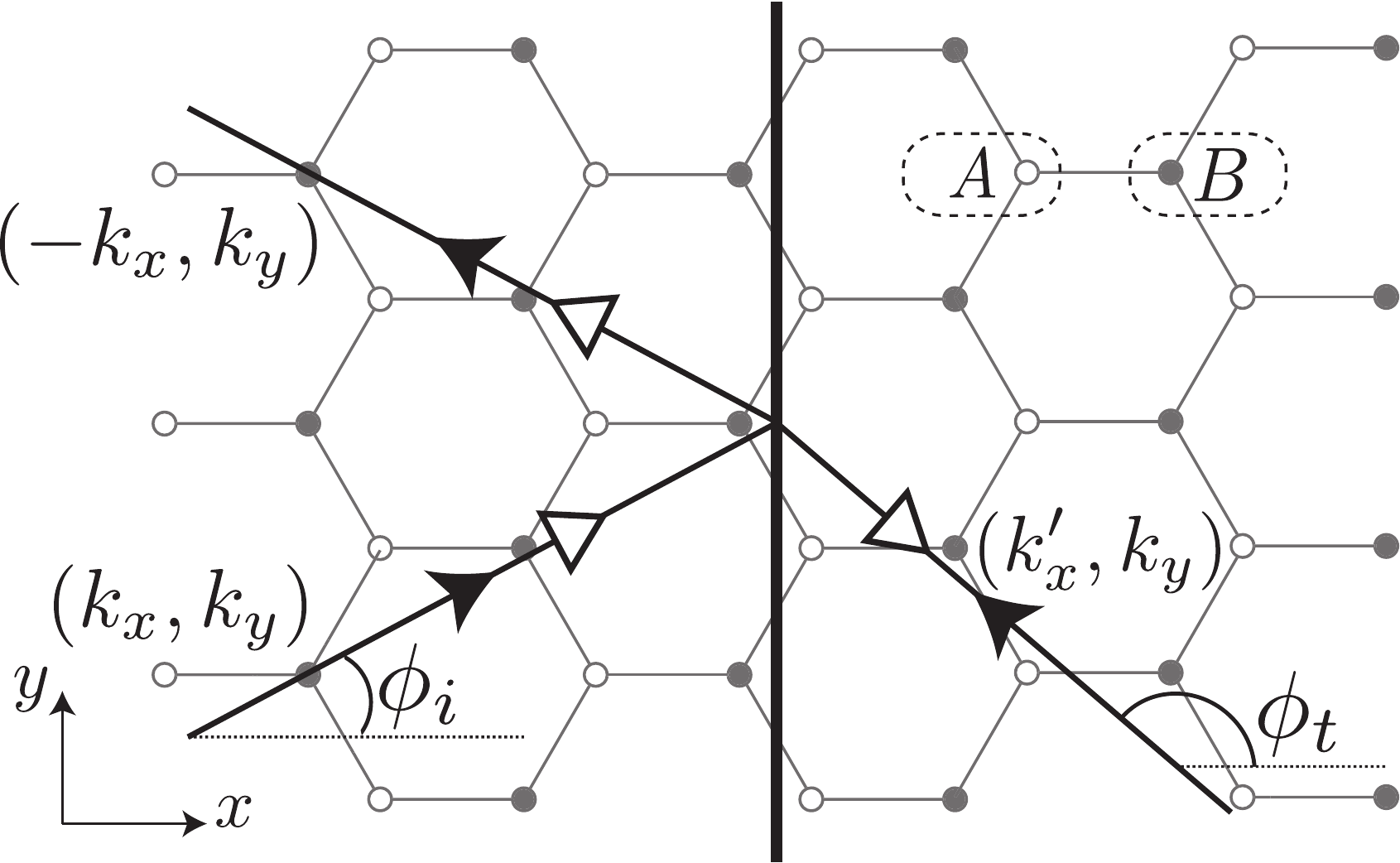}
\caption{Incident, reflected, and transmitted waves in a honeycomb lattice with a step. The thick vertical line represents the step edge; the lattice sites on the right of the thick vertical line have an additional potential energy $V$ compared to those on the left of the line.
The direction of the full arrows indicates the direction of the momentum measured from a Dirac point, while the open ones indicate the group velocity. The parameters chosen in this figure correspond to a {\em negative refraction} case.}
\label{honeycomb}
\end{center}
\end{figure}

We first briefly review the basic concepts of the Klein tunneling in conservative systems without losses. We consider a honeycomb lattice oriented as in Fig.~\ref{honeycomb} with a uniform tunneling amplitude $J$ between neighboring sites and a sharp potential step of height $V$, where the lattice sites at $x \ge 0$ have a higher energy than those at $x < 0$.

A beam of particles is incident from $x < 0$, and is partially reflected and partially transmitted at the edge of the step at $x = 0$, as described in Fig.~\ref{honeycomb}.
In both $x < 0$ and $x > 0$ regions the energy dispersion of the particles shows linear crossings at momenta called the Dirac points.
The Dirac velocity, which is the group velocity of a particle around a Dirac point, is $v_D = 3aJ/2$, where the lattice spacing is $a$ and we have set $\hbar = 1$. The zero of the energy is chosen at the Dirac point in the $x < 0$ region, while the Dirac point in the $x>0$ is displaced in energy by the potential step $V$.

To describe the wave function on the honeycomb lattice, we use a tight-binding model on the A,B sublattice basis and we assume that the relevant wave vectors are within the linear dispersion region in the vicinity of the Dirac points.
In an infinite and spatially homogeneous honeycomb lattice, the particle wave function is characterized by the crystal momentum $(k_x, k_y)$, which is measured with respect to the momentum at the Dirac point $\mathbf{K} = (K_x, K_y)$, and its angle $\phi$ defined through $k_x + ik_y = e^{i\phi} k$ with $k=({k_x^2 + k_y^2})^{1/2}$. For sufficiently small $k$, the energy is equal to $v_D\,k$. 

In the A,B sublattice basis, the wave function can be written in the following spinor form~\cite{CastroNeto2009}:
\begin{align}
	\psi (x,y)
	=
	e^{ik_x x + ik_y y + i\mathbf{K}\cdot\mathbf{r}}
	\begin{pmatrix}
	1 \\ i e^{i(\phi - K_x a - k_x a)}
	\end{pmatrix}. \label{wavefo}
\end{align}
This equation should be understood such that if one wants to calculate the wave function on a lattice site at position $\mathbf{r}$ on the A sublattice, one takes the first component of the spinor and multiplies it by a Wannier function localized around the site. Similarly, if one wants to calculate the wave function on the B sublattice, one takes the second component. Throughout this manuscript, we assume that the Wannier functions are sharply localized at the lattice sites.
Our gauge choice for the spinor wave function in (\ref{wavefo}) follows from our choice of the orientation of the honeycomb lattice and of the unit cell, where a B lattice site is displaced from an A lattice site by a distance $a$ in the $x$ direction.
While it is perhaps more customary in the graphene literature to use a spinor wave function of the form $(1, e^{i\phi})^T$ that differs from ours by an appropriate gauge transformation to the Hamiltonian~\cite{CastroNeto2009}, our choice (\ref{wavefo}) appears to be more convenient to compare with numerical simulations or actual experiments in photonic systems where we have a direct access to the spinor structure of the wave function.

\begin{figure}[htbp]
\begin{center}
\includegraphics[width=8.0cm]{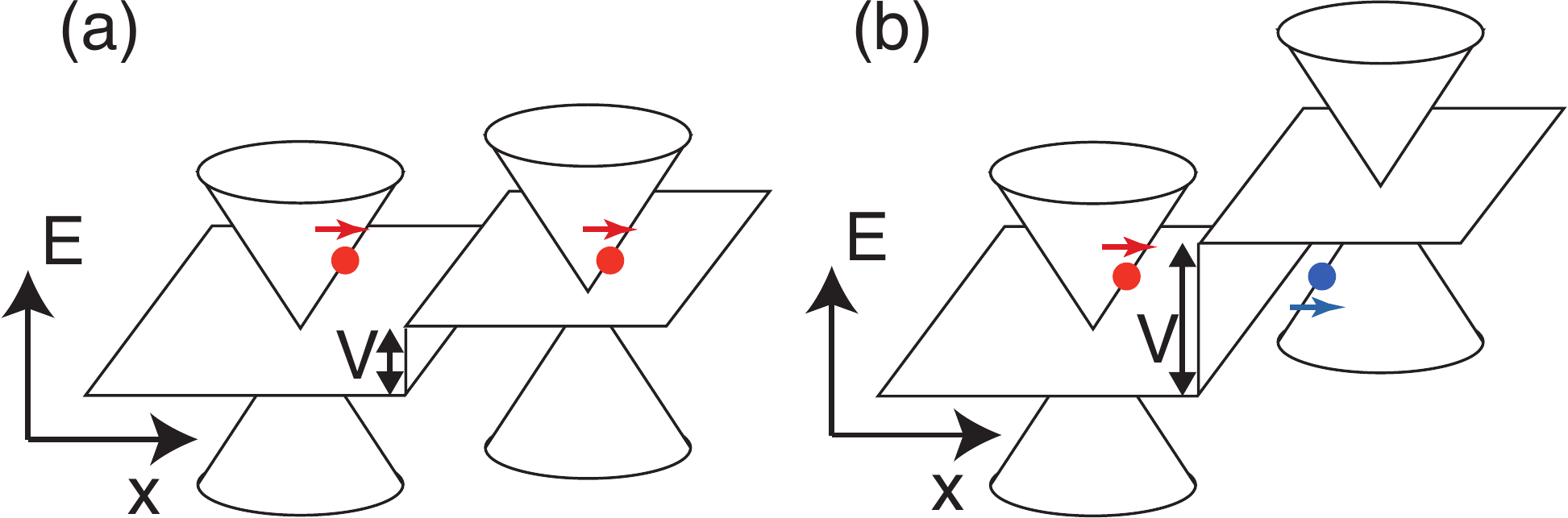}
\caption{Schematic illustration of the Klein tunneling. The cones represent the Dirac dispersion before ($x < 0$) and after ($x > 0$) the step. (a) When the step height is smaller than the energy of the incident particle. The transmitted particle is in the upper band and therefore the transmission is particlelike. (b) When the step height is larger than the energy of the incident particle. The transmitted particle is in the lower band and therefore the transmission is holelike.}
\label{schematic}
\end{center}
\end{figure}

In the following we focus our attention to the case of a beam of particles incident on the potential step with a positive energy and a given momentum. Depending on whether the step height $V$ is lower or higher than the incident energy, the transmitted beam will be ``particle-like'' or ``hole-like,'' as schematically illustrated in Fig.~\ref{schematic}.  Under the assumption that intervalley scattering processes by the potential step are negligible, we can write a general wave function in the $x < 0$ ($\psi_l$) and $x > 0$ ($\psi_r$) regions with a probability current from left to right of the step as~\cite{Beenakker2008, Allain2011}
\begin{align}
	\psi_l (x,y)
	&=
	e^{ik_x x + ik_y y + i\mathbf{K}\cdot\mathbf{r}}
	\begin{pmatrix}
	1 \\ i e^{i(\phi_i - K_x a - k_x a)}
	\end{pmatrix}
	\notag \\
	&\hspace{1cm}
	+
	r
	e^{-ik_x x + ik_y y + i\mathbf{K}\cdot\mathbf{r}}
	\begin{pmatrix}
	1 \\ ie^{i(\pi-\phi_i - K_x a + k_x a)}
	\end{pmatrix},
	\notag \\
	\psi_r (x,y)
	&=
	t e^{ik_x^\prime x + ik_y y + i\mathbf{K}\cdot\mathbf{r}}
	\begin{pmatrix}
	1 \\ -ie^{i(\phi_t - K_x a - k_x^\prime a)}
	\end{pmatrix}, \label{wavef}
\end{align}
where $(k_x, k_y)$, $(-k_x, k_y)$, and $(k_x^\prime, k_y)$ are the momenta of incident, reflected, and transmitted waves, respectively, measured with respect to a Dirac point $\mathbf{K} = (K_x, K_y)$.
More details on the validity of neglecting the intervalley scattering for a sharp step are discussed in Sec.~\ref{sec:inter}.

Translational invariance along $y$ guarantees conservation of $k_y$. The longitudinal $k_x^\prime$ is determined by energy conservation plus the condition that the group velocity of the transmitted beam must be in the positive $x$ direction. The sign of $k_x^\prime$ will depend on whether the transmission is particle- or hole-like; see Fig.~\ref{schematic}. The angles $\phi_{i,t}$ of the incident and transmitted beams are defined through $k_x + ik_y = e^{i\phi_i} ({k_x^2 + k_y^2})^{1/2}$ and $k_x^\prime + ik_y = e^{i\phi_t}({k_x^{\prime 2} + k_y^2})^{1/2}$. Depending on the particle or hole character of the transmitted beam, the $y$ component of the group velocity will have the same or the opposite sign as compared to the incident beam; the latter case (sketched in Fig.~\ref{honeycomb}) goes under the name of {\em negative refraction}.

The reflectivity $R$ and the transmittivity $T$ are related to the reflection and transmission coefficients $r$ and $t$ and the angles $\phi_i$ and $\phi_t$ through~\cite{Allain2011}
\begin{align}
	R &= \frac{\cos \phi_i |r|^2}{\cos \phi_i |r|^2 + |\cos \phi_t||t|^2},
	\notag \\
	T &= \frac{|\cos \phi_t||t|^2}{\cos \phi_i |r|^2 + |\cos \phi_t||t|^2}.
	\label{randt}
\end{align}
Neglecting intervalley coupling, one can determine $r$ and $t$ by requiring that the wave function is continuous at $x = 0$, which yields the following theoretical predictions:
\begin{align}
	R &= 
	\frac{1 - \cos (\phi_i - \phi_t)}{1 + \cos (\phi_i + \phi_t)},
	&
	T &=
	\frac{2\cos \phi_i \cos \phi_t}{1 + \cos (\phi_i + \phi_t)}, \label{randttheorypart}
\end{align}
for the particlelike transmission when $V < \omega$, and
\begin{align}
	R &= 
	\frac{1 + \cos (\phi_i - \phi_t)}{1 - \cos (\phi_i + \phi_t)},
	&
	T &=
	-\frac{2\cos \phi_i \cos \phi_t}{1 - \cos (\phi_i + \phi_t)}, \label{randttheoryhole}
\end{align}
for the holelike transmission when $V > \omega$,
where $\omega$ is the incident energy. 

In particular, when the beam is normally incident ($\phi_i = 0$ and $\phi_t = 0$ or $\pi$), the transmittivity is exactly one independently of the potential step height, resulting in a perfect transmission without any backscattering.
This peculiar tunneling effect where a particle transmits through a potential which is higher than its energy is called the Klein tunneling, in analogy with the Klein paradox for relativistic particles~\cite{Klein1929, Baym}. In the context of particles in a honeycomb lattice this phenomenon arises as a consequence of the pseudo-spin (chirality) conservation on both sides of the step~\cite{Allain2011}.

In the context of graphene physics, Klein tunneling has been studied in the case of a finite-width barrier~\cite{Katsnelson2006}, rather than a step, and an oscillatory behavior of the transmission rate was predicted and observed as one changes the incident angle~\cite{Stander2009, Young2009}.
In the following, we show that in photonic systems one can directly observe the Klein tunneling for a single step, a situation closer to the original argument of Klein, via the analysis of both the angle dependence and the step height dependence of the conductivity.

\section{Photonic graphene}
\label{photgra}

\subsection{Theoretical model}

We consider a photonic graphene, such as the one realized in~\cite{Jacqmin2014}, in which photons have a finite lifetime.
By continuous wave resonant pumping of the system, a steady-state configuration of photons is reached, whose real- and momentum-space distributions can be experimentally measured by detecting the near-field and far-field emissions, respectively. We restrict to the linear regime in which photons do not interact.

We consider a tight-binding Hamiltonian in which the annihilation operators of photons at position $\mathbf{r}$ in A and B sublattices are denoted by $\hat{a}_{\mathbf{r}}$ and $\hat{b}_{\mathbf{r}}$, respectively. Note that $\mathbf{r}$ takes only discrete values on lattice sites.
Under monochromatic coherent pumping, the expectation values of the operators in the Heisenberg representation, $a_{\mathbf{r}}(t) = \langle \hat{a}_{\mathbf{r}}(t)\rangle$ and $b_{\mathbf{r}}(t) = \langle \hat{b}_{\mathbf{r}}(t)\rangle$, evolve according to the pump frequency $\omega$ as $a_{\mathbf{r}}(t) = a_{\mathbf{r}} e^{-i\omega t}$ and $b_{\mathbf{r}}(t) = b_{\mathbf{r}} e^{-i\omega t}$.

In this paper, we consider a uniform loss for all sites at a rate $\gamma$.
Then, the steady-state configuration of the photon fields $a_{\mathbf{r}}$ and $b_{\mathbf{r}}$ is obtained by solving the following linear equations~\cite{Ozawa2014}:
\begin{align}
	&f^a_{\mathbf{r}}
	=
	\left(
	\omega + i\gamma - V_{\mathbf{r}}
	\right)
	a_{\mathbf{r}}
	\notag \\
	&\hspace{2cm}
	+
	J
	\left(
	b_{\mathbf{r}+\boldsymbol{\delta}_1} + b_{\mathbf{r}+\boldsymbol{\delta}_2} + b_{\mathbf{r}+\boldsymbol{\delta}_3}
	\right),
	\notag \\
	&f^b_{\mathbf{r}}
	=
	\left(
	\omega + i\gamma - V_{\mathbf{r}}
	\right)
	b_{\mathbf{r}}
	\notag \\
	&\hspace{2cm}
	+
	J
	\left(
	a_{\mathbf{r}-\boldsymbol{\delta}_1} + a_{\mathbf{r}-\boldsymbol{\delta}_2} + a_{\mathbf{r}-\boldsymbol{\delta}_3}
	\right), \label{lineareq}
\end{align}
where $f_{\mathbf{r}}^a$ and $f_{\mathbf{r}}^b$ are the spatial amplitude profile of the pump field acting on A and B sublattices, respectively, and $V_{\mathbf{r}}$ is the step height which is $V_{\mathbf{r}} = V>0$ at $x \ge 0$ and $V_{\mathbf{r}} = 0$ at $x < 0$. The vectors $\boldsymbol{\delta}_1 \equiv (a,0)$, $\boldsymbol{\delta}_2 \equiv (-a/2,-a\sqrt{3}/2)$, and $\boldsymbol{\delta}_3 \equiv (-a/2,a\sqrt{3}/2)$ connect the nearest neighbors of a honeycomb lattice.

This model is not limited to the region of linear dispersion around the Dirac points; it reproduces the whole band structure when considering cylindrically symmetric photonic modes at each lattice site coupled to their nearest neighbors. This is exactly the case of lattices of microwave resonators~\cite{Bittner:2010, Bellec:2013}, single-mode waveguides~\cite{Rechtsman:2013PRL} or micropillar polaritons~\cite{Jacqmin2014}. In the case of polaritons, photoluminescence experiments have directly shown the complete band structure of the lattice, including the linear dispersion close to the Dirac points, in agreement with Eq.~(\ref{lineareq}). In the following calculations we will make use of the complete Hamiltonian~(\ref{lineareq}) though our focus will mostly be on modes close to the Dirac points.

One of the assets of photonic systems is that a step potential $V$ can be easily implemented in lattices of resonators or waveguides~\cite{Bittner:2010, Bellec:2013, Rechtsman:2013PRL, Jacqmin2014}. This can be done by increasing or reducing the diameter of the resonators at a given region of the sample. In this way, the confinement energy of the photonic mode is modified resulting in a different onsite energy. As the resonators have a size comparable to the wavelength of the considered photon---on the order of micrometers in the case of polariton microcavities---the onsite energy can be easily controlled using standard lithographic techniques. Here we will concentrate on a step potential in which the onsite energy is abruptly changed on the scale of one lattice site.

\subsection{Coherent pumping and imaging}
\label{sec:cohpump}

We now describe the pumping scheme which creates the steady-state configuration of photon fields acting as the incident wave for the Klein tunneling.
We consider the situation where the system is coherently pumped by a Gaussian field with spatial profile described by $f(\mathbf{r}) = e^{i\mathbf{k}_c \cdot \mathbf{r} - (\mathbf{r} - \mathbf{r}_0)^2/2\sigma^2}$, where $\mathbf{r}_0$ is the center of the pump, $\mathbf{k}_c$ is the central momentum, and $\sigma$ is the spatial width of the pump.

In the steady-state configuration of photons in the absence of a step, all states with energy $\omega$ and momentum covered by the pump can in principle be excited, but the weight of excitation of each state depends on the complex overlap between the A and B sublattice components in the spinor wave function~(\ref{wavefo}) and the pump profile. The two sublattices can in fact interfere constructively or destructively with the pump depending on the momentum.

\begin{figure}[htbp]
\begin{center}
\includegraphics[width = 0.49 \textwidth]{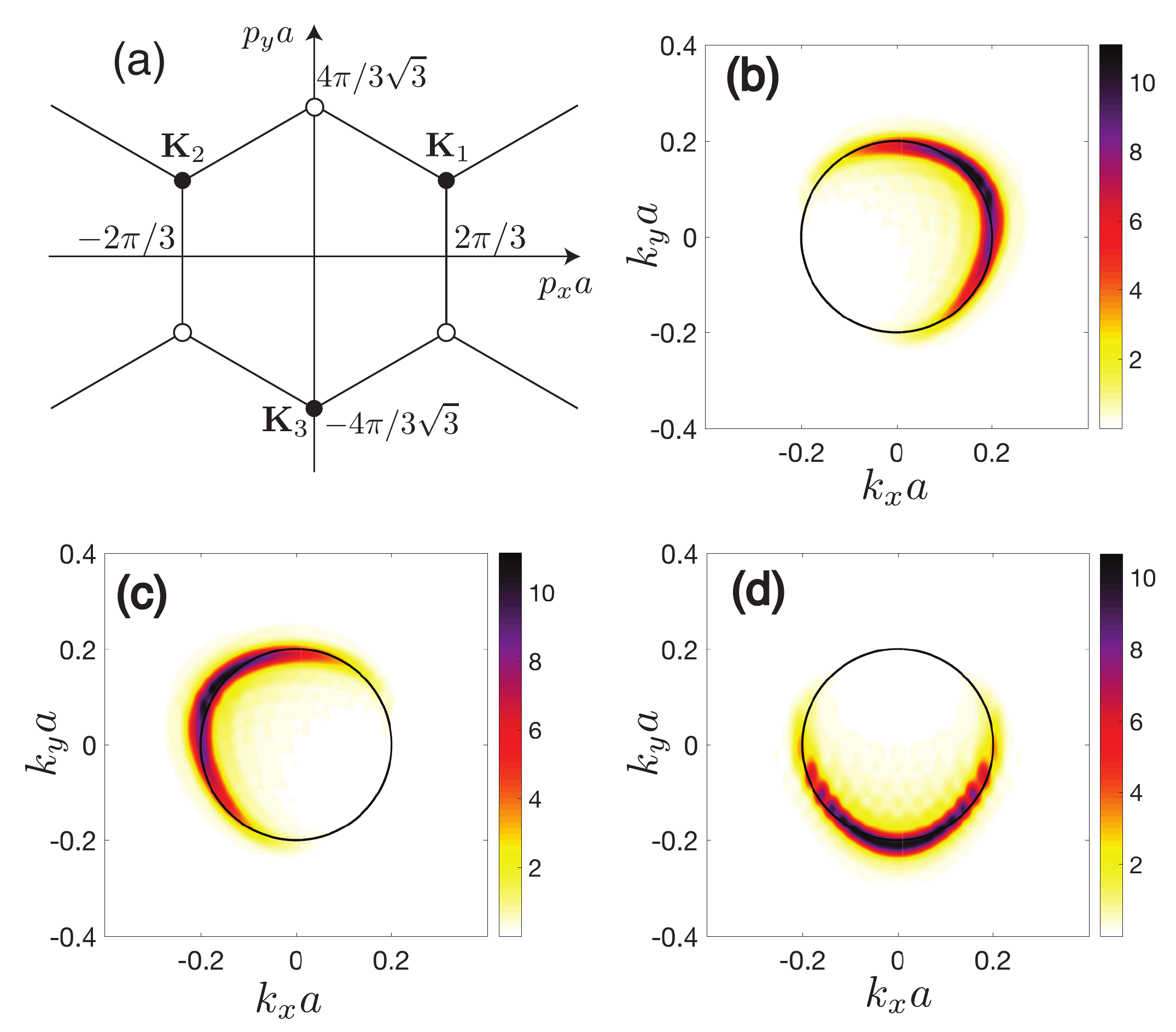}
\caption{(a) First Brillouin zone of a honeycomb lattice denoted by a hexagon in momentum space, whose vertices are the Dirac points. The solid circles, $\mathbf{K}_1$, $\mathbf{K}_2$, and $\mathbf{K}_3$, are the Dirac points in the same valley which we mainly explore in this paper. The hollow circles are the Dirac points in the other valley. (b)-(d) The momentum space distribution of the steady-state emission of photons when the spatially tightly focused pump fields are concentrated around the different Dirac points $\mathbf{K}_1$, $\mathbf{K}_2$, and $\mathbf{K}_3$ and the emission is detected around the same respective Dirac points. The black circles are the iso-energy surfaces corresponding to the pump energy. The width of the pumping field is $\sigma = 5 a$.}
\label{interf}
\end{center}
\end{figure}

To better understand the interference effect between A and B sublattices, in Fig.~\ref{interf}, we plot the momentum distribution of the steady-state photon fields in the absence of a step for the pumping field centered at three different Dirac points $\mathbf{K}_{1,2,3}$ and a detection window centered around the same Dirac points.
Here we consider an ideal case of a large sample size ($200 \times 200$ unit cells) and small loss $\gamma/J = 0.02$, to better illustrate our principle; we later discuss in Sec.\ref{sec:smaller} the effect of having a small sample with larger loss to consider more realistic situations. The finite size of the lattice is accounted for in our calculations by imposing an amplitude of the wave function equal to zero out of the considered area. The frequency of the pump beam is $\omega = 0.3J$, which corresponds to the momentum $k = 0.2/a$ measured from Dirac points.

Figure~\ref{interf} shows that, on iso-energy surfaces, the emission is concentrated around different angles $\phi$ for a pump located around different Dirac points.
Here we observe different emissions around different Dirac points, even though these Dirac points are equivalent in the sense that they are related by reciprocal lattice vectors. The reason for the different behavior around different Dirac points is that our lattice has two lattice sites per unit cell which gives rise to a geometrical structure factor effect. This effect is analogous to that in solid-state electron systems, where Bragg reflection peaks associated with reciprocal lattice vectors can have intensity variations due to the geometrical structure factor~\cite{AshcroftMermin}. It has been observed in angle-resolved photoemission of graphite~\cite{Shirley1995} and graphene~\cite{Bostwick2007}, and in the photoluminescence of a honeycomb lattice of micropillars~\cite{Jacqmin2014}.

This angle-dependent emission pattern can be quantitatively explained by the structure of the spinor wave function in Eq.~(\ref{wavefo})~\cite{Jacqmin2014}.
To understand the pattern, it is convenient to regard the emission as a result of two separate processes; (i) a given pump beam excites a particular mode, and (ii) the excited mode emits light. Both processes have angle dependence, and the resulting emission pattern is the product of the two processes.
The first process can be understood from the overlap between the pump beam localized around a Dirac point, e.g., $\mathbf{K}_1$, and the spinor (\ref{wavefo}). From the spinor structure, we see that the mode at momentum $(k_x, k_y)$ and angle $\phi$ in the vicinity of the Dirac point $\mathbf{K}_1$ is excited with the strength of $|1 + ie^{i(\phi - K_{1,x} a)}|^2$.
To understand the second process, let us assume that a mode at momentum $(k_x, k_y)$ around a Dirac point $\mathbf{K}_1$ is excited with a unit strength. Then, if the detection is performed around the same Dirac point $\mathbf{K}_1$, this mode emits light with the intensity of $|1 + ie^{i(\phi - K_{1,x} a)}|^2$, which gives the structure factor for the emitted light. In the end, as a result of the two processes, the observed emitted intensity from the momentum $(k_x, k_y)$ for a pump located around $\mathbf{K}_1$ is approximately $|1 + ie^{i(\phi - K_{1,x} a)}|^4$. Instead if the detection is performed around a different Dirac point $\mathbf{K}_2$, the first process does not change if we use the same pump, but the second process gives the intensity of $|1 + ie^{i(\phi - K_{2,x} a)}|^2$, and the observed emitted intensity is $|1 + ie^{i(\phi - K_{1,x} a)}|^2 |1 + ie^{i(\phi - K_{2,x} a)}|^2$. Below, we discuss how to exploit this angle dependence of the sublattice interference to selectively excite an incident beam which mainly propagates toward the potential step.

From the momentum distribution of the steady-state in the presence of a step, we want to extract the information on the reflection coefficient $|r|^2$ and the transmission coefficient $|t|^2$ defined in Eq.~(\ref{wavef}). When doing this, one needs to keep in mind that $r$ and $t$ are the coefficients in the sublattice basis.
As one performs the Fourier transform of both sublattices at, for example, $x > 0$, and takes the intensity of the momentum component corresponding to $\mathbf{K}_1 + (k_x^\prime, k_y)$, one obtains a value proportional to $|t (1 -e^{i(\phi_t - k_x^\prime - K_{1,x} a)})|^2\simeq |t (1 -e^{i(\phi_t - K_{1,x} a)})|^2$, not just $|t|^2$.
Note that although these expressions, derived from Eq.~(\ref{wavef}), are defined for a conservative system, they also apply to the case of weak and homogeneous losses in a sufficiently large lattice.

While the angle dependence may have some utility to selectively focus on a specific range of momenta, one can avoid the angle-dependent factor by summing the momentum components of the emitted intensity around three adjacent Dirac points $\mathbf{K}_1$, $\mathbf{K}_2$, and $\mathbf{K}_3$ in Fig.~\ref{interf}(a), or, equivalently, around three Dirac points which are separated by $4\pi/3a$ in the $k_x^\prime$ direction, which gives
\begin{align}
	&|t (1 -e^{i(\phi_t - k_x^\prime a - K_{1,x} a)})|^2 + |t (1 -e^{i(\phi_t - k_x^\prime a - 4\pi/3 - K_{1,x} a)})|^2
	\notag \\
	&+ |t (1 -e^{i(\phi_t - k_x^\prime a - 8\pi/3 - K_{1,x} a)})|^2 =
	6|t|^2. \label{angularfactor}
\end{align}
An analogous result holds for the reflection coefficient $|r|^2$. We have assumed here that the Wannier function is well localized in space so that its Fourier transform has the same weight for all momentum components.

\section{Klein tunneling in photonics}
\label{sec:schemes}

We are now ready to present our proposal to study Klein tunneling in driven-dissipative systems. In the next two subsections we will discuss two possible schemes, based on, respectively, a spatially tightly focused pump beam and a spatially wide pump beam. In the former case, a $k_y$-selective detection is sufficient to reconstruct the full angle dependence of Klein tunneling using a single pump spot. In the latter case, an incident beam with a well-defined wave vector $\mathbf{k}$ is excited, which allows one to also visualize the negative refraction effect. 

Throughout this section, we focus on a rather idealized case of a large lattice and a small loss rate. For all considered cases, we will show that the quantities extracted from the numerical simulation are in good agreement with the general theory of Klein tunneling. In the next section (Sec.~\ref{sec:smaller}), we shall see how this conclusion survives when more realistic parameters from actual experiments are used. 

\subsection{Spatially focused beam}
\label{spfb}

In this subsection, we consider a beam focused in space ($\sigma = 5 a$), and analyze tunneling with a step $V = 0.4J$. We take the loss to be $\gamma/J = 0.02$ throughout this section.
We use the Dirac point $\mathbf{K}_1 \equiv (2\pi/3a, 2\pi/3\sqrt{3}a)$ to be the central momentum of the pump field and the pump frequency of $\omega = 0.3J$, which corresponds to the momentum $k = 0.2/a$ from the Dirac point.
We place the center of the pump $\mathbf{r}_0$ to be in the middle between the lower-left corner of the sample and the center of the edge of the step at $x = 0$.
As one sees from Fig.~\ref{interf}(b), the momentum distribution of the no-step configuration is strong in $k_x > 0$, thus pumping the field at $x < 0$ mainly excites an incident beam that propagates towards the potential step.

\begin{figure}[htbp]
\begin{center}
\includegraphics[width = 0.24\textwidth]{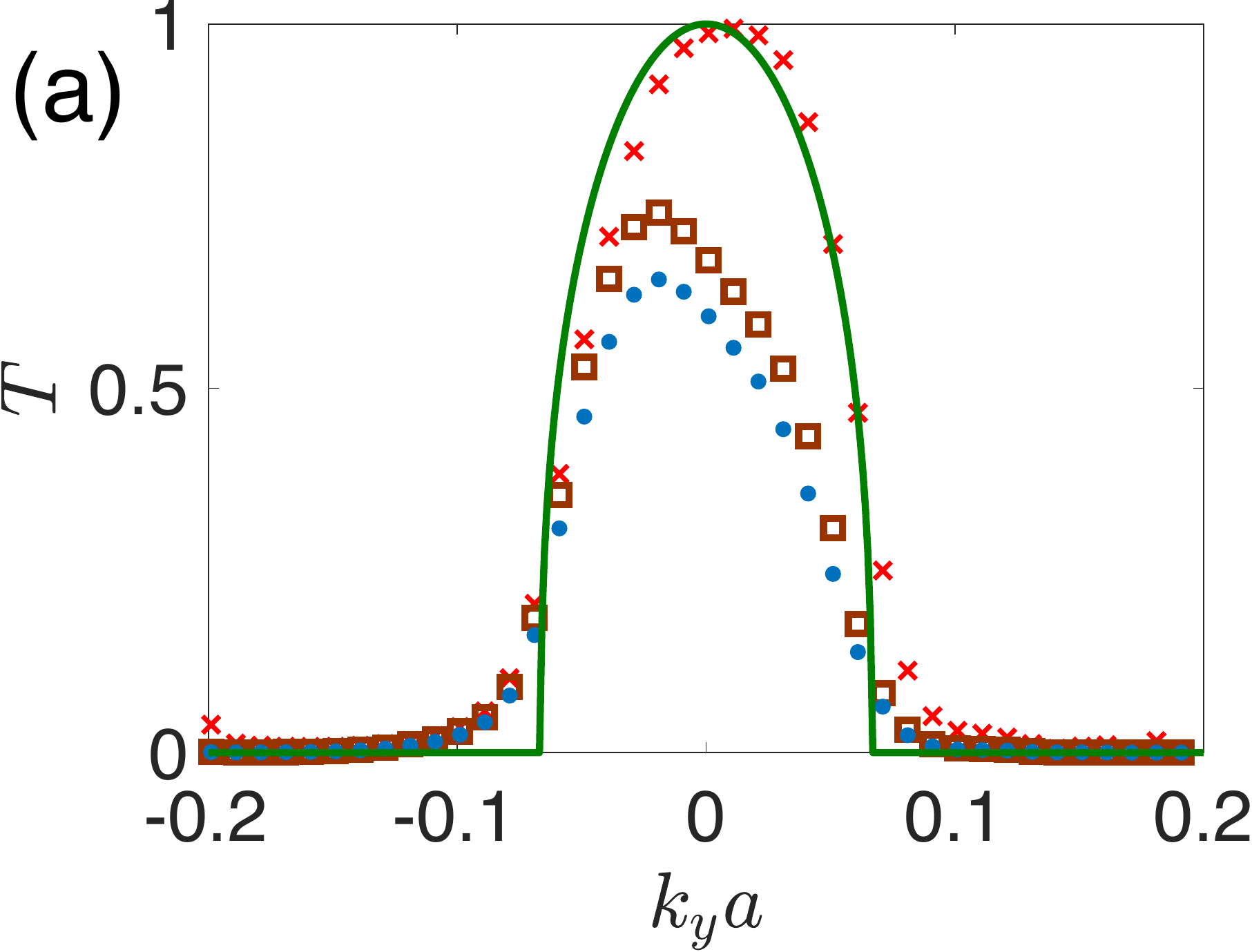}
\includegraphics[width = 0.23\textwidth]{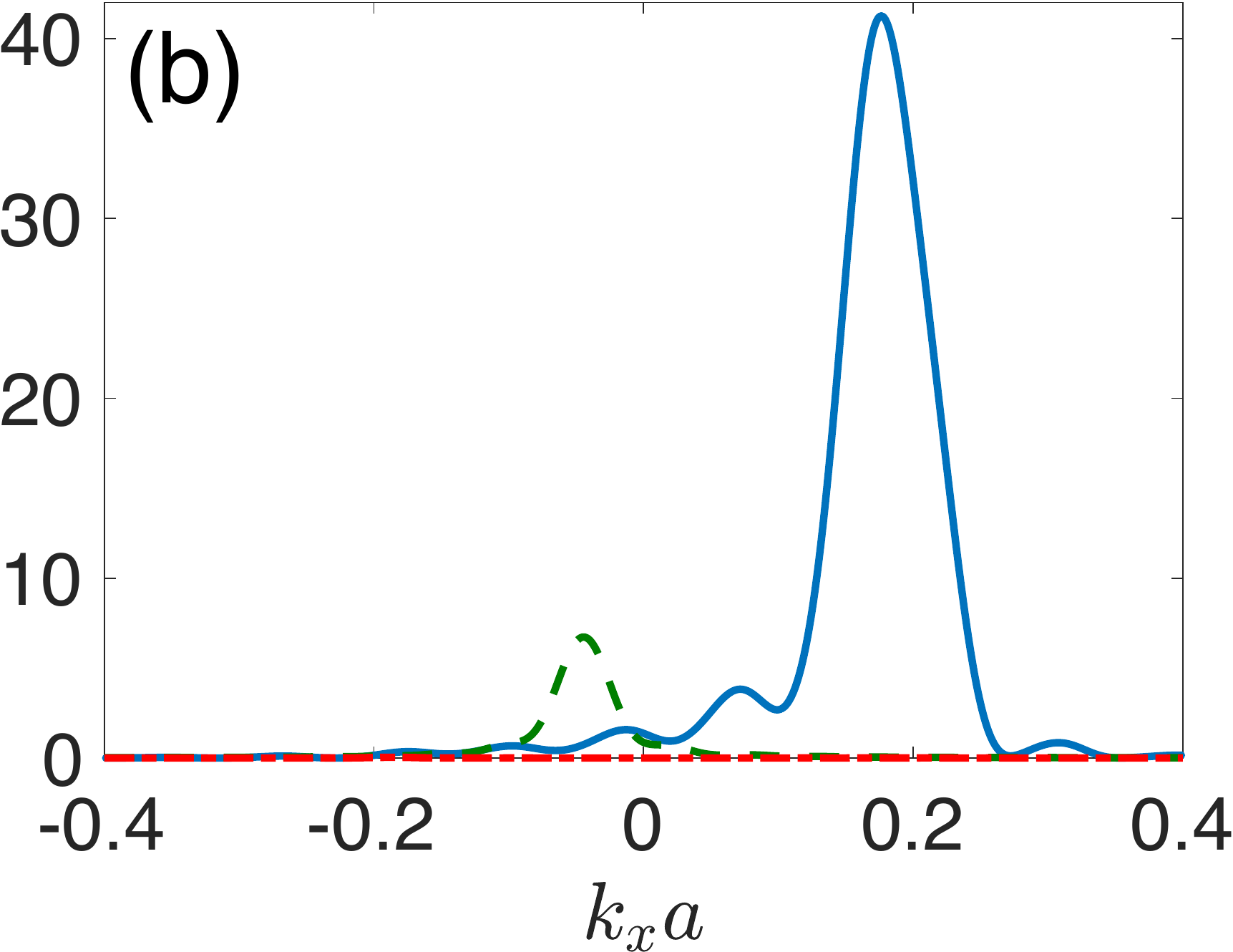}
\includegraphics[width = 0.23\textwidth]{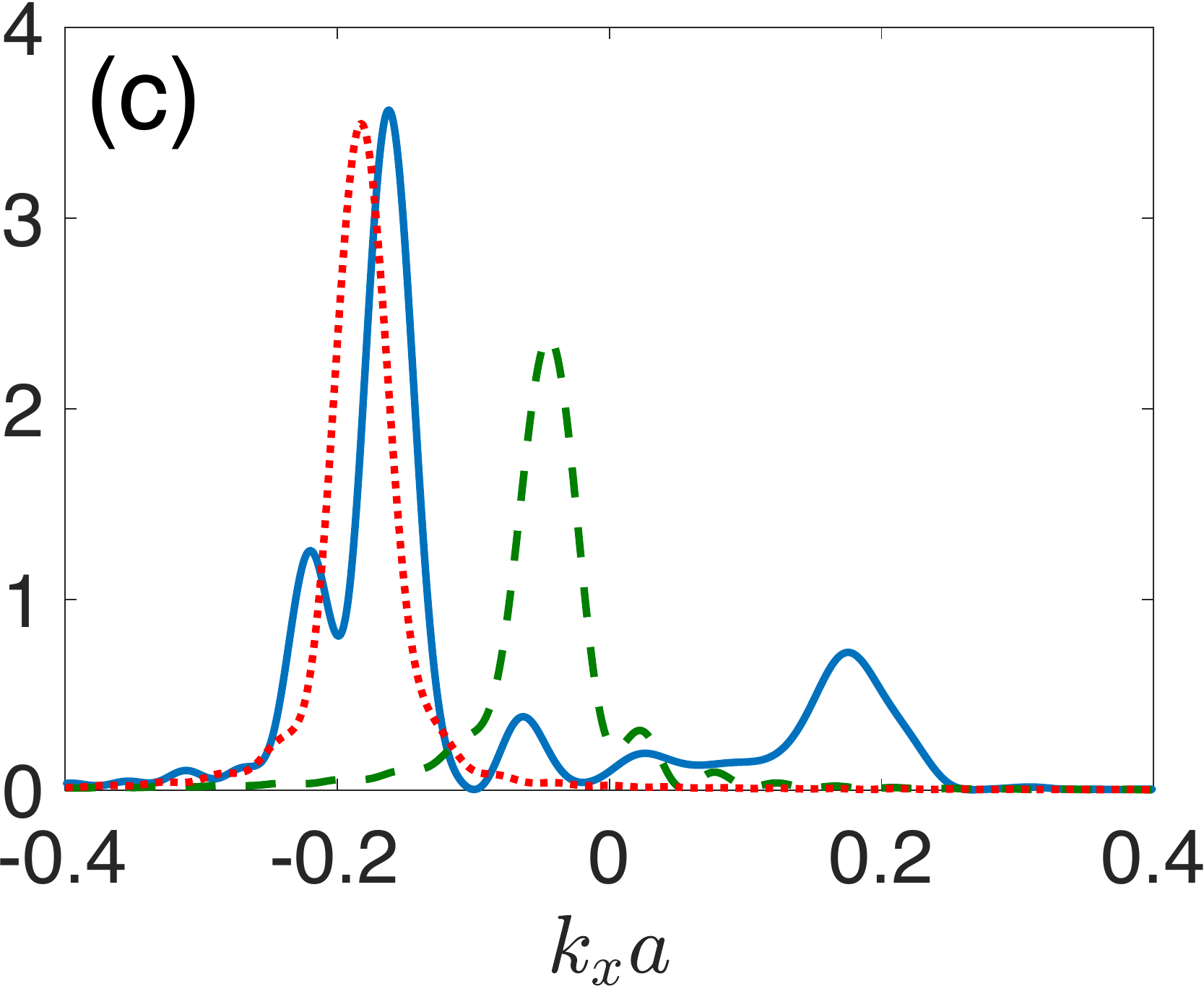}
\includegraphics[width = 0.23\textwidth]{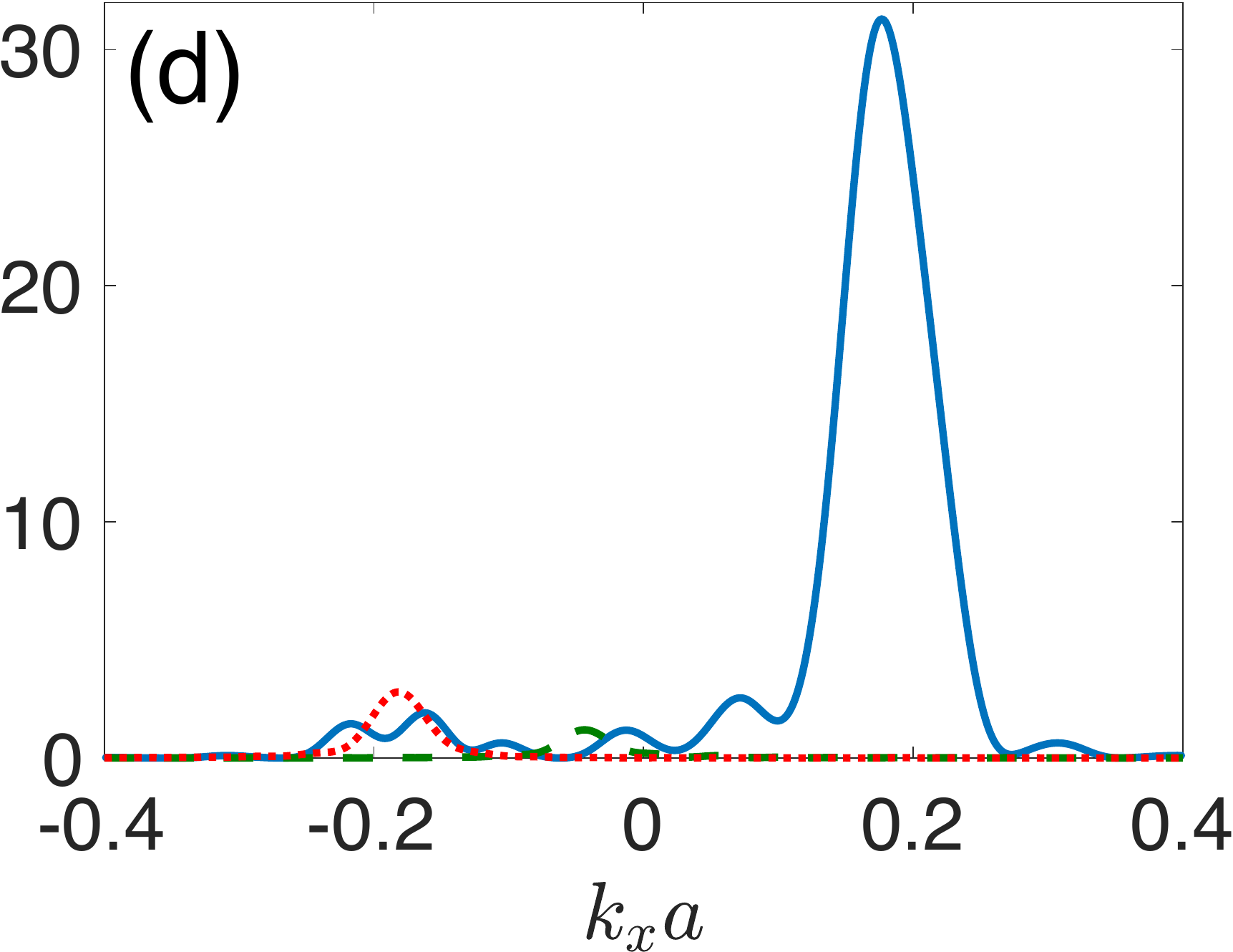}
\caption{
(a) Transmittivity $T$ as a function of $k_y$. The line is a theoretical curve from (\ref{randttheoryhole}). The dots (blue) are calculated from the simulation with the reflection coefficient $|r|^2$ estimated by summing the signals at $x < 0$ around $\mathbf{K}_1$, $\mathbf{K}_2$, and $\mathbf{K}_3$, the brown squares are the transmittivity calculated by only using signals emitted around $\mathbf{K}_2$, and the red crosses are the simulation where the reflection coefficient $|r|^2$ is estimated from the signal at $x < 0$ after subtracting the no-step distribution to isolate the reflection signal.
(b)-(d) Cut of the momentum distribution for $k_y = -0.05/a$ as a function of $k_x$ around $\mathbf{K}_1$, $\mathbf{K}_2$, and $\mathbf{K}_3$, respectively. The vertical axis is dimensionless photon counts, and we use the same scale for (b)-(d). The solid (blue) lines are the spatial Fourier transform of the field in the $x < 0$ region, and the dashed (green) lines are the spatial Fourier transform of the field in the $x > 0$ region. The dotted (red) lines are the spatial Fourier transform of the $x < 0$ region after subtracting the no-step distribution.}
\label{subtracted}
\end{center}
\end{figure}

In Figs.~\ref{subtracted}(b)-(d), we plot a cut of the momentum distribution for a given $k_y = -0.05/a$ as a function of $k_x$, where $(k_x, k_y)$ is measured with respect to the three Dirac points $\mathbf{K}_1$, $\mathbf{K}_2$, and $\mathbf{K}_3$, respectively. The momentum distribution of the $x < 0$ region is plotted in solid (blue) lines, which contains the information of both the incident and reflected waves, and the momentum distribution of the $x > 0$ region is plotted in dashed (green) lines, which represents only the transmitted wave.
From (\ref{wavef}), in the case of no loss with a plane wave with a single wave vector as an incident wave, one expects that the momentum space signal would be two sharp peaks at $x < 0$, corresponding to the incident and reflected waves, and one sharp peak at $x > 0$, corresponding to the transmitted wave. In our system, as we see in Figs.~\ref{subtracted}(b)-(d), the peaks are broadened due to the loss, and also have some structures due to the presence of the sample edges.

The different structures observed in Figs.~\ref{subtracted}(b)-(d) can be understood in terms of the sublattice interference effect discussed also in the previous section. We observe that the incident wave ($k_x > 0$ component of the solid lines) is well visible in Figs.~\ref{subtracted}(b) and \ref{subtracted}(d) but is strongly suppressed in Fig.~\ref{subtracted}(c). This is because, as seen from Fig.~\ref{interf}(c), the emission at $k_x > 0$ is small around $\mathbf{K}_2$.
We also observe that the $k_x < 0$ part of the signal of Fig.~\ref{subtracted}(c) is not a single peak but shows a visible modulation on top of the peak. This is because the pump beam generates weak excitations also in $k_x < 0$ modes which travel backward and reflect at the edge of the sample, thus producing the interference effects.

We estimate the reflection coefficient $|r|^2$ by integrating the $k_x < 0$ part of the detected signal of the $x < 0$ region (solid lines) and summing over the three Dirac points $\mathbf{K}_1$, $\mathbf{K}_2$, and $\mathbf{K}_3$, as explained in the end of Sec.~\ref{sec:cohpump}. We note that this method generally overestimates the reflection because the $k_x < 0$ component of the incident beam, although small due to the choice of our pump beam, is not excluded in the integral.
The transmission coefficient $|t|^2$ can be estimated by integrating the signal of $x > 0$ (dashed lines). To find the reflection and transmission rates $R$ and $T$ from~(\ref{randt}), we also need to know $\phi_i$ and $\phi_t$.
The incident angle $\phi_i$ is given for a fixed value of $k_y$, and the transmission angle is determined by finding the average value of the transmission signal in the momentum space, which we call $k_x^t$, and then through the relation $k_x^t + i k_y = \sqrt{k_x^{t 2} + k_y^2} e^{i\phi_t}$.

In Fig.~\ref{subtracted}(a), we plot the transmission rate $T$ as a function of $k_y$. 
The dots (blue) are calculated from the numerical simulation, and the line is the analytical prediction of Eq.~(\ref{randttheoryhole}). We observe that the transmittivity is quite underestimated in the numerical simulation. 
This is because the signal at $x < 0$ to estimate the reflection coefficient contains contributions from incident waves as well. The overestimation of the reflection wave is larger at $k_y > 0$ than at $k_y < 0$ because the incident wave, whose profile is similar to Fig.~\ref{interf}(b), has more overlap with the $k_y > 0$ region than with the $k_y < 0$ region at $k_x < 0$ and thus overestimates the reflection at $k_y > 0$.

One can improve the estimate by pumping around $\mathbf{K}_1$ and detecting only around $\mathbf{K}_2$, taking advantage of the fact that the incident wave is very suppressed around $\mathbf{K}_2$ as we discussed above. In order to calculate the transmittivity using the signal only around $\mathbf{K}_2$, upon integrating the signal at $x < 0$, one needs to take into account the angular factor due to the spinor structure correctly to estimate $|r|^2$ and $|t|^2$, because there is no cancellation as in (\ref{angularfactor}). The transmittivity thus calculated is plotted as brown squares in Fig.~\ref{subtracted}(a). We see an improvement in the estimate of the transmittivity. The advantage of this method is that one needs to only measure around one Dirac point.

Finally, a very efficient but experimentally challenging way to isolate the reflected wave in the $x<0$ region is to subtract the real-space field amplitude distribution without a step from the one with a step before performing the Fourier transform; the resulting momentum distribution is plotted in dotted (red) lines in Figs.~\ref{subtracted}(b)-(d).
It shows a single peak at $k_x < 0$ as we would expect for photons with nonzero $k_y$ component; we recall that backscattering is only forbidden for exactly normal incidence onto the step.
The transmission calculated using this procedure is displayed as red crosses in Fig.~\ref{subtracted}(a), which shows a very good agreement with the theoretical prediction.

\subsection{Intervalley scattering}
\label{sec:inter}

At this point, it is worth discussing the importance of the intervalley scattering in our system. For the vertical step we consider (see Fig.~\ref{honeycomb}), the intervalley scattering is kinematically not allowed due to the simple momentum and energy conservation. Namely, as one can see from Fig.~\ref{interf}(a), there is no state in the other valley conserving the energy and the momentum $k_y$ in the vertical direction. 
On the other hand, if the step is aligned horizontally, there is a state in the other valley conserving the energy and the momentum $k_x$ in the horizontal direction. Therefore, if we were to use a sharp horizontal step, we would have a non-negligible amount of the intervalley scattering. Such an intervalley scattering would be suppressed if one uses not a sharp but a smooth step~\cite{Allain2011}.

\begin{figure}[htbp]
\begin{center}
\includegraphics[width = 0.49 \textwidth]{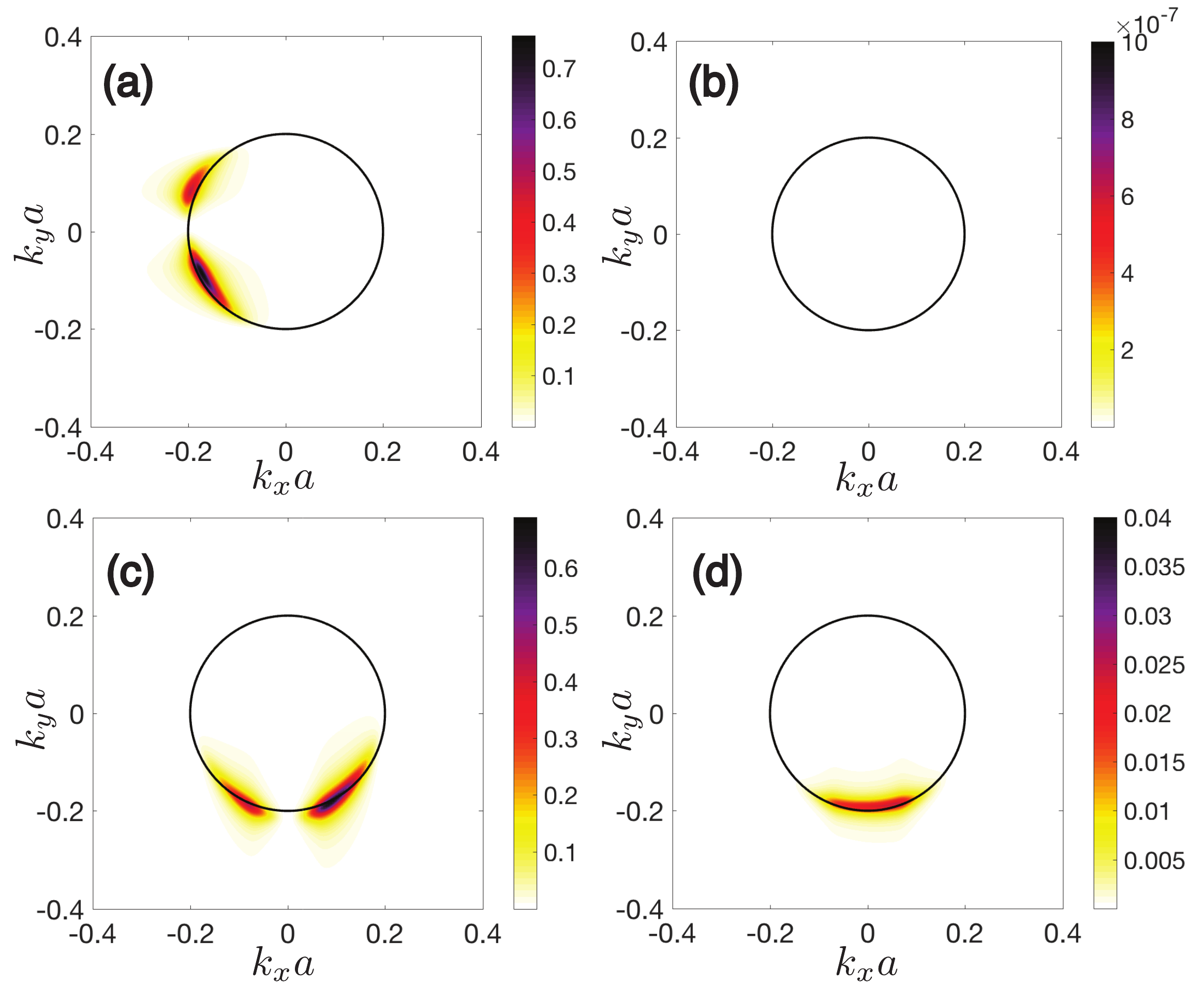}
\caption{Momentum-space distribution of the steady-state emission of photons around Dirac points calculated for the reflected signal, where the incident signal is subtracted to isolate the reflected signal. (a) and (b) When the step is vertical located at $x = 0$, and (c) and (d) when the step is horizontal located at $y=0$. In (a) and (c), the emission around the equivalent Dirac points $\mathbf{K}_1$, $\mathbf{K}_2$, and $\mathbf{K}_3$ are summed, whereas in (b) and (d), the emission around the Dirac points in the other valley is summed. The simulation is done on a very large lattice with a size of $1000 \times 1000$. The central momentum of the pump beam is at $\mathbf{K}_1$ with the spot size of $\sigma = 5 a$. The pump is located at 100 lattice sites away from the center of the system in (a) and (b) the horizontal direction, and (c) and (d) the vertical direction.}
\label{intervalley}
\end{center}
\end{figure}

This orientation dependence of the intervalley scattering for a sharp step is confirmed by the numerical simulation as shown in Fig.~\ref{intervalley}.
In Figs.~\ref{intervalley}(a) and \ref{intervalley}(b), we plot the momentum-space distribution of the steady-state emission of photons of the reflected beam when the step is vertical. In order to isolate the reflection signal, we only look at the $x < 0$ part of the signal and we subtracted, as before, the steady-state photon amplitude without a step from that with a step. Figure~\ref{intervalley}(a) is the sum of the emission around the Dirac points $\mathbf{K}_1$, $\mathbf{K}_2$, and $\mathbf{K}_3$, and Fig.~\ref{intervalley}(b) is the sum of the emission around the Dirac points of the other valley. We see essentially no emission from the other valley. In Figs.~\ref{intervalley}(c) and \ref{intervalley}(d), we plot the case when the step is horizontal. Fig.~\ref{intervalley}(c) is the sum around $\mathbf{K}_1$, $\mathbf{K}_2$, and $\mathbf{K}_3$, and Figure~\ref{intervalley}(d) is the sum around the Dirac points of the other valley. We see that the emission around the other valley is now significant.
Thus, the vertical orientation of the step we use in this paper is ideal for studying the Klein tunneling without intervalley scattering. 
We note that the absence of the reflected signal around $k_y a = 0$ for Fig.~\ref{intervalley}(a) and around $k_x a = 0$ for Fig.~\ref{intervalley}(c) is the manifestation of the perfect transmission for the normal incidence in the Klein tunneling.

In reality, if we use a finite-size system, the intervalley scattering can also occur at the top and bottom edges of the finite-size lattice, which are oriented horizontally. In the following analysis, the scattering at the edges is not siginificant for the large system size we use in Sec.~\ref{sec:schemes}. However, for small systems in Sec.~\ref{sec:smaller}, the scattering at the edges is one of the major sources of discrepancy from the numerical simulation and the analytical prediction of the Klein tunneling.

\subsection{Wide beam}

We now proceed with the discussion of an alternative method to study Klein tunneling effects which is also able to elucidate the negative refraction effect. The pump field is taken to have a wide Gaussian profile in real space with $\sigma = 20 a$ and a momentum distribution well localized around the central momentum $\mathbf{k}_c = \mathbf{K}_1 + k (\cos 45^\circ, \sin 45^\circ)$ with $k = 0.2/a$. The pump frequency is set to $\omega = 0.3J$ on resonance with the Dirac dispersion at the central momentum of the pump.

\begin{figure}[htbp]
\begin{center}
\includegraphics[width = 0.49 \textwidth]{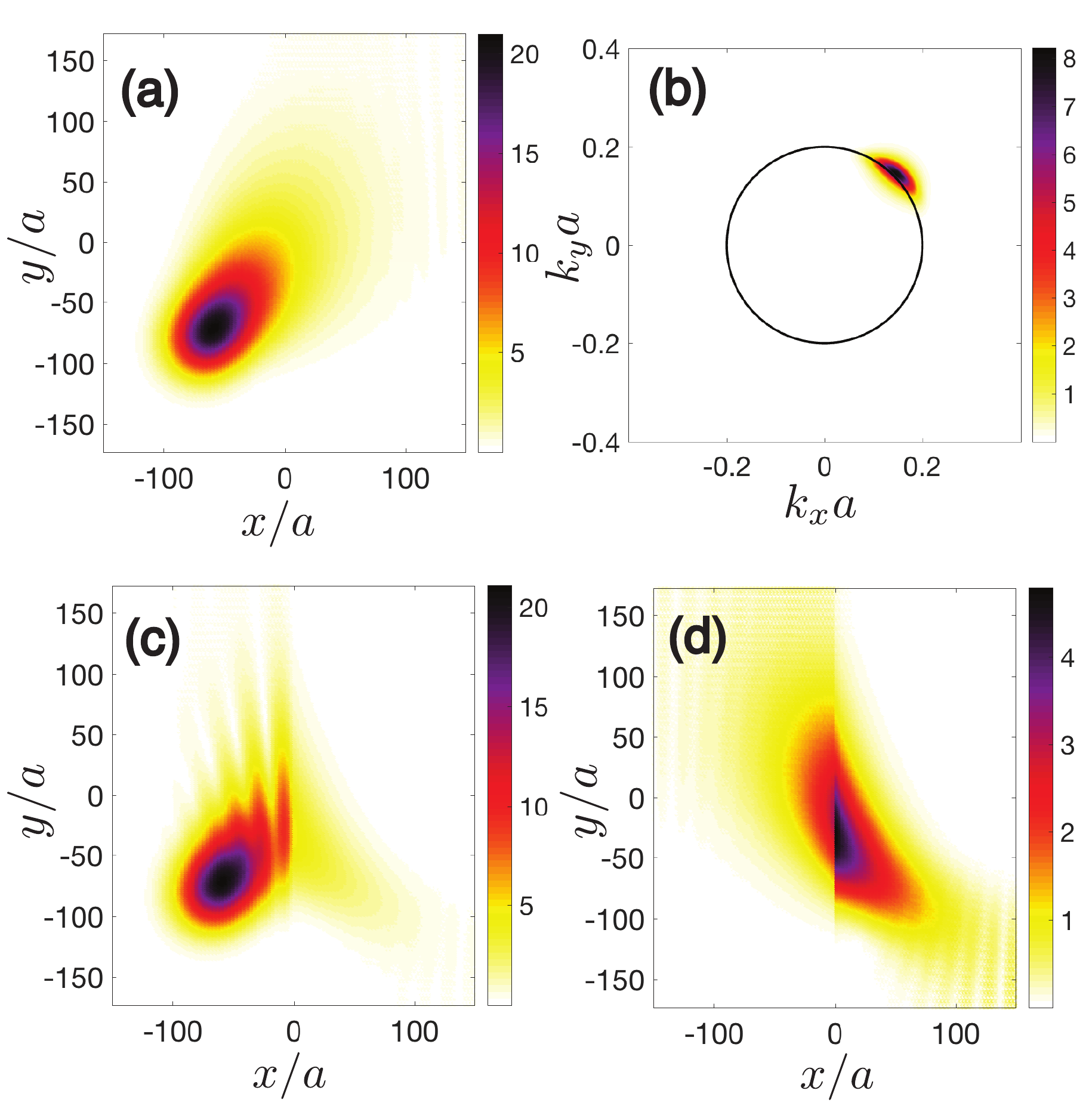}
\caption{Field intensity profiles under a spatially wide pump. Real-space (a) and momentum-space around $\mathbf{K}_1$ (b) distributions of the photon emission for no-step potential. (c) Real-space profile for a high potential step $V=0.6J>\omega$. (d) Real space profile after subtracting the no-step amplitude in the $x<0$ region. In all panels, the pump energy is $\omega=0.3J$ and the loss is $\gamma = 0.02J$. The other parameters are given in the text.}
\label{typical}
\end{center}
\end{figure}

In Figs.~\ref{typical}(a) and \ref{typical}(b), we plot the steady-state field profile in real and momentum spaces in the absence of a step. One can see that photons propagate in real space at an angle of $45^\circ$ and the incident beam is well localized in momentum space. In Fig.~\ref{typical}(c), we plot the steady-state spatial configuration of photons in the presence of a step with $V = 0.6J$, where one sees the existence of the reflection and the transmission.
The transmission through the step is directed downward, which shows the negative refraction characteristic of the Klein tunneling in graphene for a hole-like transmission when the potential step is higher than the incident energy~\cite{Allain2011}. The usual refraction is recovered in the opposite case of a potential step lower than the incident energy (not shown).

To better visualize the reflected and transmitted fields, in Fig.~\ref{typical}(d) we show the field amplitude after subtracting in the $x<0$ region the one for the no-barrier case [Fig.~\ref{typical}(a)].
Another salient feature we observe from Fig.~\ref{typical}(d) is that the centers of the reflected and transmitted waves at the step edge are shifted. This is an analog of the Goos-H\"anchen effect known for total internal reflection in optical systems~\cite{Goos1947}. However, the value of the shift obtained in the numerical simulations does not fit the conventional theory of the Goos-H\"anchen effect in solid-state or photonic graphene~\cite{Beenakker2009, Grosche2016}. Therefore, a more elaborate theory taking into account pump and loss is necessary and will be the subject of future work.

\subsubsection{Changing the angle}

The use of an incident beam localized in momentum provides a convenient method to study the angular dependence of the Klein tunneling effect. To perform this study we fix the step height at $V = 0.4J$ and consider an incident beam well localized in momentum space (spatial width $\sigma = 40 a$) around $\mathbf{k}_c = \mathbf{K}_1 + k (\cos \phi_i, \sin \phi_i)$ with $k = 0.2/a$ and $\omega = 0.3J$. We vary the incidence angle $\phi_i$.

For a given value of $\phi_i$, we Fourier transform the steady-state amplitude to obtain the momentum-space distribution in the $(k_x, k_y)$ plane.
We analyze the reflection and transmission waves by looking at the momentum distribution as a function of $k_x$ for a fixed $k_y = k \sin \phi_i$.

\begin{figure}[htbp]
\begin{center}
\includegraphics[width = 0.225\textwidth]{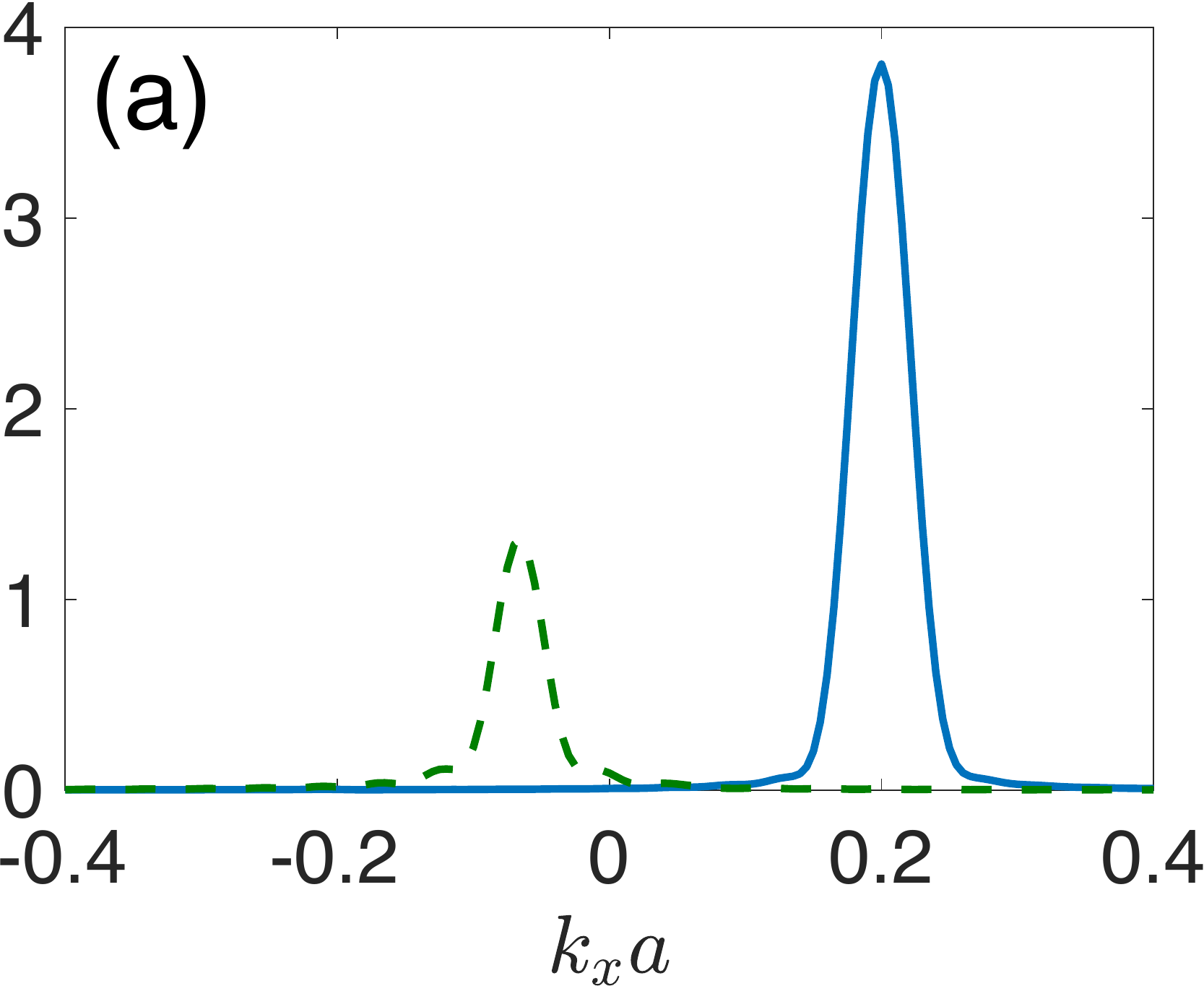}
\includegraphics[width = 0.235\textwidth]{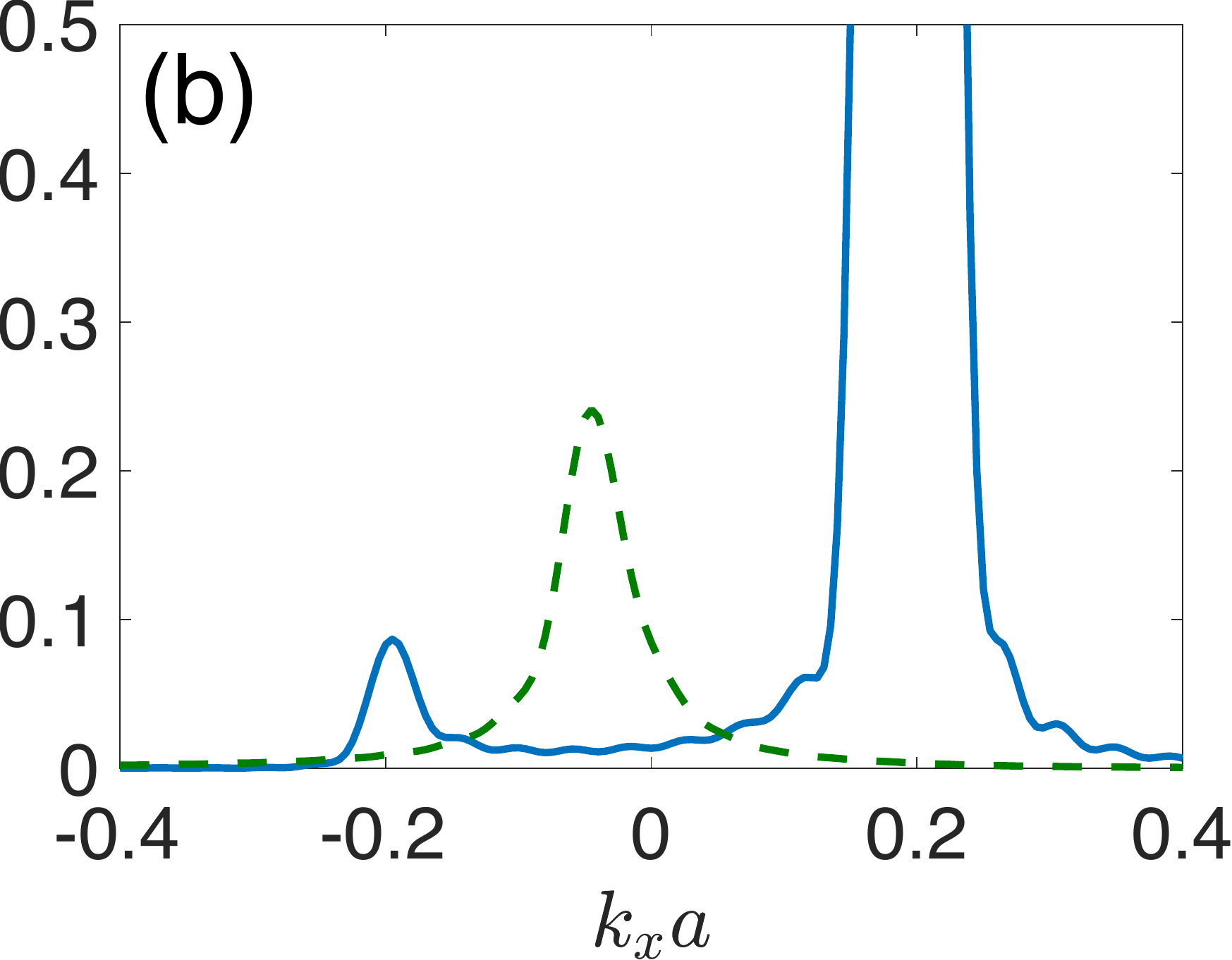}
\includegraphics[width = 0.23\textwidth]{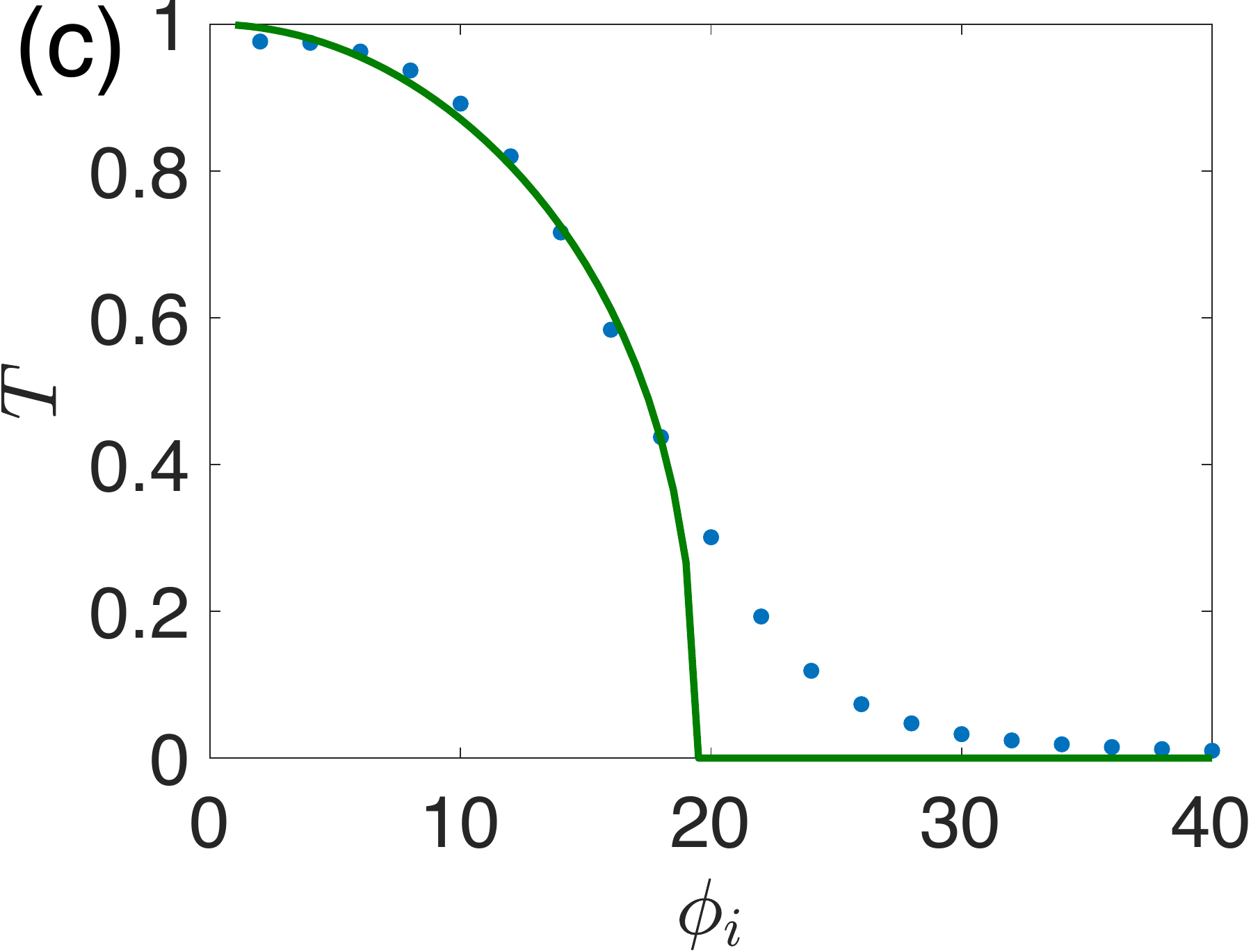}
\includegraphics[width = 0.23\textwidth]{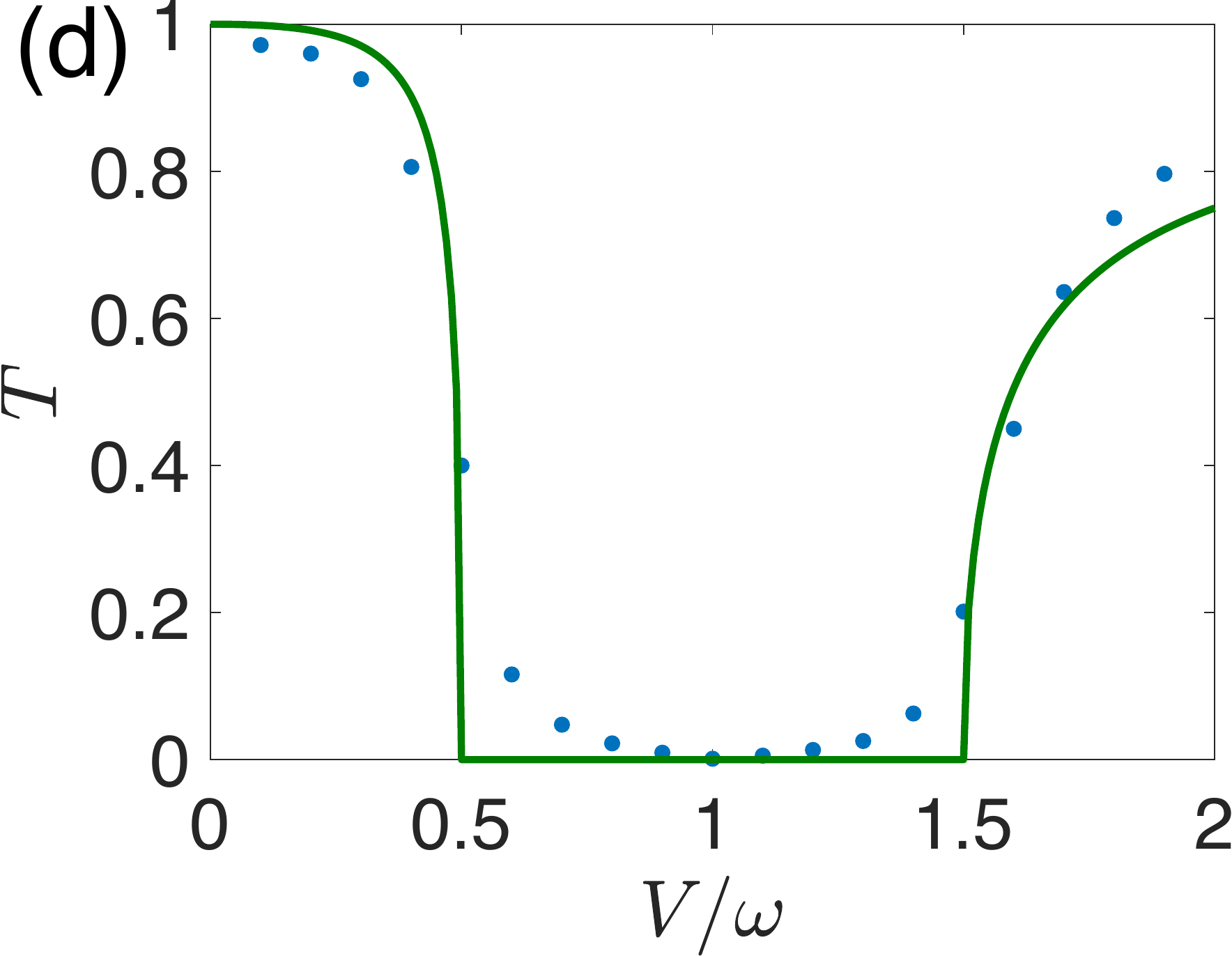}
\caption{Momentum-space distribution of the steady-state emission of photon fields for (a) $\phi_i = 0$ ($k_y = 0$) and (b) $\phi_i = 18^\circ$ [$k_y = 0.2\sin (18^\circ)/a$] as a function of $k_x$, measured from $K_{1,x}$. The vertical axes of (a) and (b) are plotted with the same scale. The solid line is the Fourier transform at $x < 0$, and the dashed line is that at $x > 0$.
The transmission rates when (c) the incident angle (in degrees) is changed for a fixed value of $V = 0.4J$ and (d) the height of the step is changed for a fixed value of $\phi_i = 30^\circ$. The dots are calculated numerically, and the solid lines are the theoretical prediction of (\ref{randttheorypart}) and (\ref{randttheoryhole}).
The calculations are done on a lattice with $200 \times 200$ unit cells with $\gamma/J = 0.02$.}
\label{figideal}
\end{center}
\end{figure}

In Fig.~\ref{figideal}(a), we show the normal incidence ($\phi_i = 0$) case.  The solid and dashed lines are the Fourier transform in the $x < 0$ and $x > 0$ regions, respectively.  As we expect from the theory of Klein tunneling, there is only one peak in the Fourier transform of $x < 0$ corresponding to the incident wave, and no peak from reflection is present.

On the other hand, in Fig.~\ref{figideal}(b), we plot the momentum distribution for a finite incidence angle $\phi_i = 18^\circ$. In this case, the signal of the incoming wave and that of the reflected wave show themselves as two separate peaks in momentum space; the larger (right) peak corresponds to the incoming wave, and the smaller (left) peak corresponds to the reflected wave.

To estimate the reflection coefficient $|r|^2$, we integrate, as before, the reflection signal at $k_x < 0$. Similarly, we estimate the transmission coefficient $|t|^2$ by integrating the transmission signal. In Fig.~\ref{figideal}(c), we plot the angle dependence of the transmission rate thus obtained.  The solid curve is the theoretical prediction from Eq.~(\ref{randttheoryhole}).

Although the qualitative agreement is generally good, there is an overestimation of transmission especially at the region beyond the critical angle $\sim 19^\circ$ where theory would predict no transmission. The observed discrepancy for angles bigger than $\sim 19^\circ$ is due to the exponentially decaying evanescent wave that is present at $x>0$ even when there can be no transmitted propagating wave. On the other hand, the well-localized momentum distribution makes the estimation of $|r|^2$ very accurate. As a result, the transmission below the critical angle is very well estimated by the numerical simulation.

\subsubsection{Changing the step height}
Next, we consider the case where the angle of the incoming wave is fixed to $30^\circ$, but the height of the step varies.
We assume that the pump beam is peaked around the momentum $\mathbf{k}_c = \mathbf{K} + k (\cos (30^\circ), \sin (30^\circ))$ with $k = 0.2/a$, and the spatial width $\sigma = 40 a$ and frequency $\omega = 0.3J$.
We vary the height of the step so that the ratio of the step height to the energy of the pump ($V/\omega$) varies from $0$ to $2$.

In Fig.~\ref{figideal}(d) we plot the transmission rate $T$ as a function of $V/\omega$.  The analytical theory (solid line) predicts a characteristic region of forbidden transmission at $0.5 \le V/\omega \le 1.5$. Below this region ($V/\omega < 0.5$), the transmission is particlelike, and above this region ($V/\omega > 1.5$), the transmission is holelike~\cite{Beenakker2008,Allain2011}. This pronounced dip is quantitatively well reproduced by the numerical simulation (dots).

\section{Smaller sample with larger loss}
\label{sec:smaller}

\begin{figure}[htbp]
\begin{center}
\includegraphics[width = 0.225\textwidth]{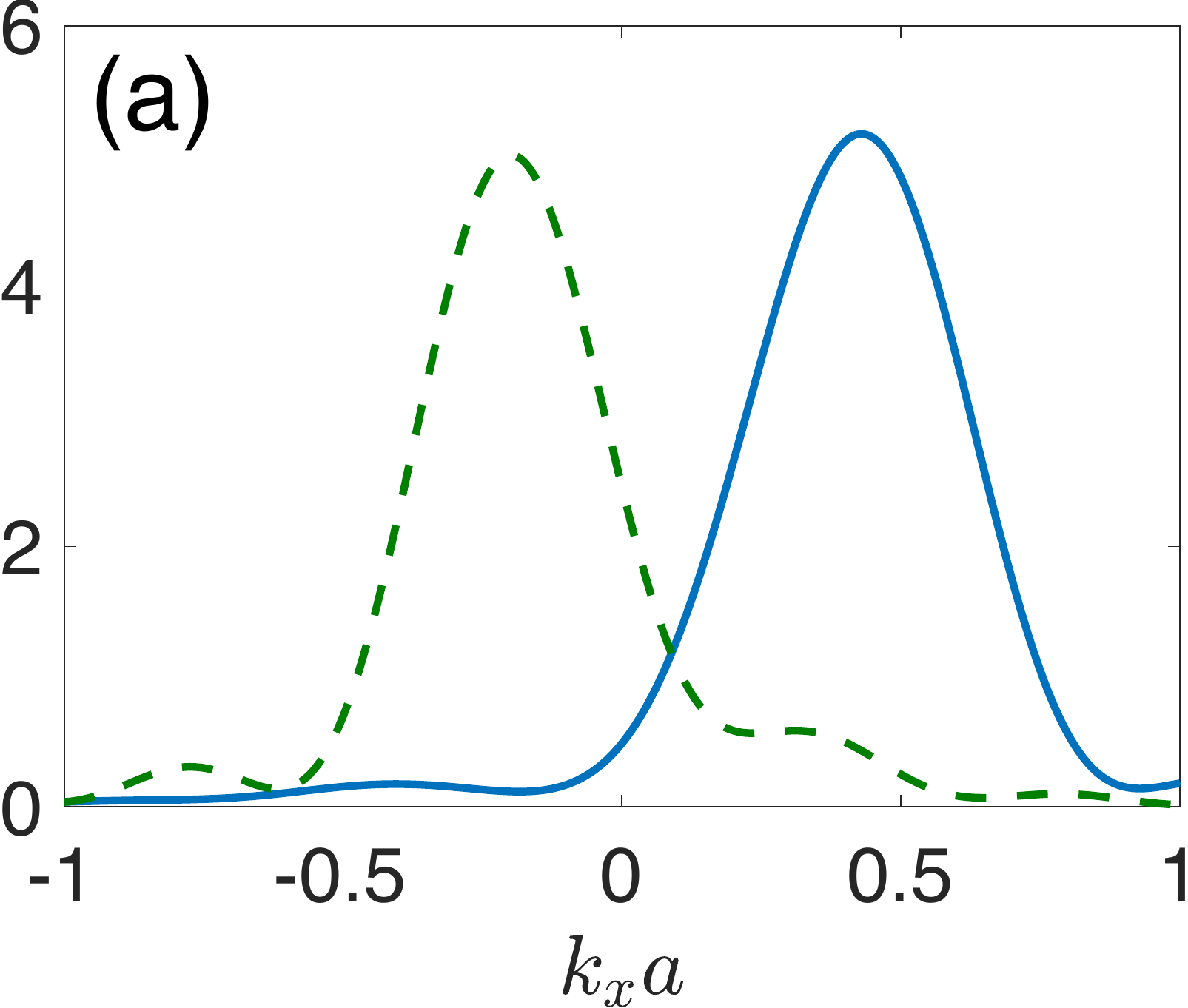}
\includegraphics[width = 0.235\textwidth]{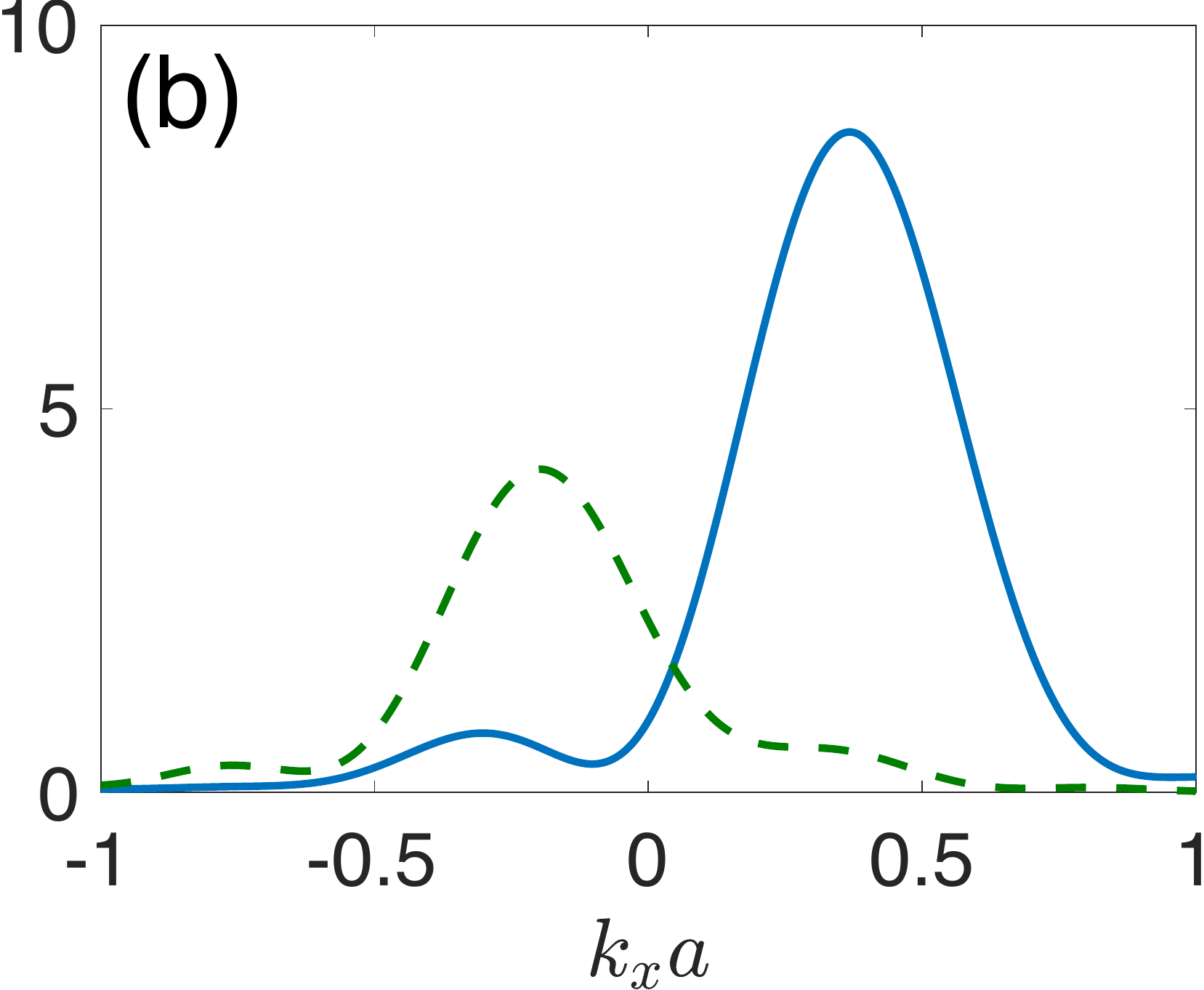}
\includegraphics[width = 0.23\textwidth]{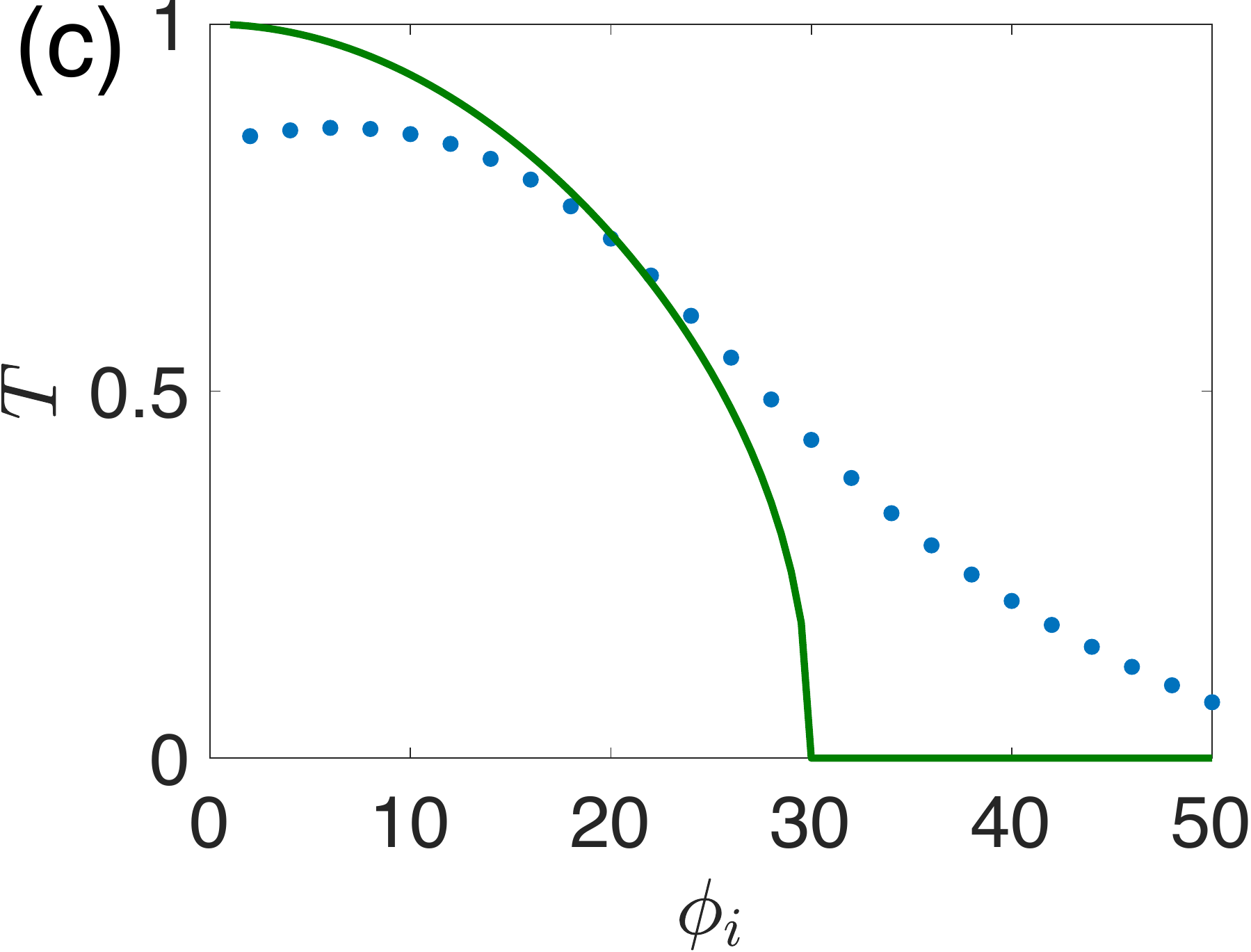}
\includegraphics[width = 0.23\textwidth]{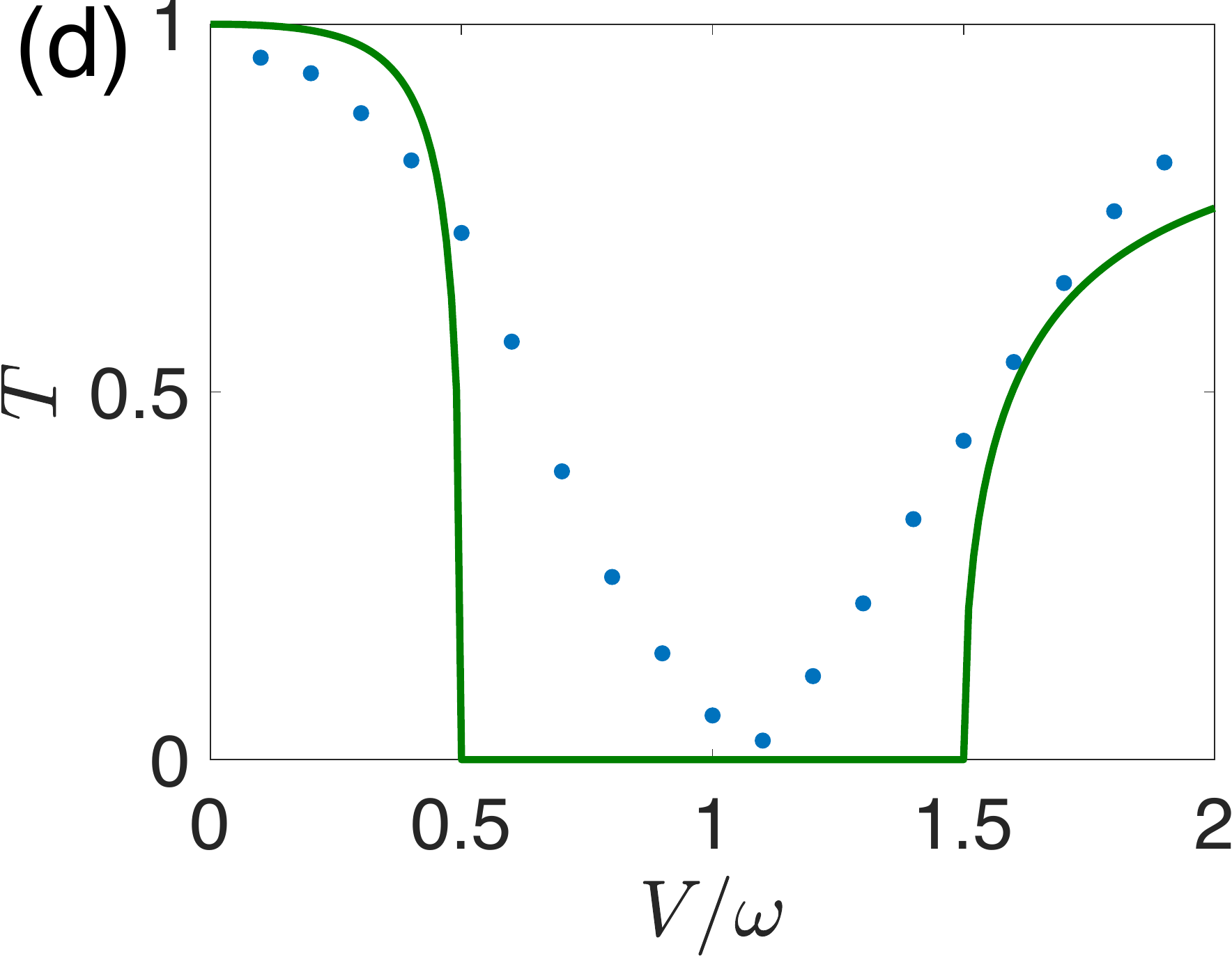}
\caption{(a) Cuts of the steady-state momentum distribution for (a) $\phi_i = 0^\circ$ and (b) $\phi_i = 18^\circ$ ($k_y = 0.4\sin (18^\circ)/a$) as a function of $k_x$, measured from $K_{1,x}$. The same scale is used for the vertical axes for both plots. The solid lines are the spatial Fourier transforms in the $x < 0$ region, and the dashed lines are the ones in the $x > 0$ region.
The transmission rates when (c) the incident angle is changed for a fixed $V = 0.9J$ and (d) the height of the step is changed for a fixed $\phi_i = 30^\circ$. The dots are calculated numerically, and the solid lines are the analytical predictions of (\ref{randttheorypart}) and (\ref{randttheoryhole}).
The calculations are done on experimentally realistic lattice parameters from~\cite{Jacqmin2014}: a lattice of $20 \times 20$ unit cells with $\gamma/J = 0.1$.}
\label{figreal}
\end{center}
\end{figure}

So far we focused on a rather large system with a small loss to demonstrate the principle of Klein tunneling in photonic systems.
Now we consider more reasonable experimental parameters from~\cite{Jacqmin2014}: a lattice of $20 \times 20$ unit cells with $\gamma/J = 0.1$.
Note that in the experiments reported in Ref.~\cite{Jacqmin2014}, no effects of the discretization of energies due to the finite size of the lattice were observed. Nevertheless, in such a small lattice, pumping with a spatially focused beam, as done in Sec.~\ref{spfb}, is not a convenient technique. Indeed, the small spot in real space excites states with $k_x<0$, which are reflected by the sample edge located at $x<0$. This edge-reflected component interferes with both the pump beam and the signal reflected from the potential step. This multiple interference prevents a clear analysis of the Klein signal.

Therefore we focus on a pumping scheme with a spatially wide beam.
However, a difficulty in dealing with a small sample is that one cannot use a pump field too sharply localized in momentum space due to the limitation in the available space in the sample. Furthermore, in addition to the larger losses giving rise to a shorter propagation distance, they also lead to a wider range of states being excited over an energy range of $\gamma$ around the pump frequency. One way to tackle these difficulties is to use a larger value of $\omega$ to increase the relevant photon momentum, so that the momentum-space peaks can be more clearly resolved. An upper bound to the allowed $\omega$ is set by the size of the linear dispersion regime around the Dirac point.

In Fig.~\ref{figreal}(a), we plot the momentum distribution of normally incident beam ($\phi_i = 0$) using a pump with central momentum $\mathbf{k}_c = \mathbf{K}_1 + k (\cos (\phi_i), \sin (\phi_i))$ and $\sigma = 10 a$, with $k = 0.4/a$ corresponding to $\omega = 0.6J$, in a spot located halfway between the lower-left corner of the sample and the center of the edge of the step at $x = 0$ as before.
We take the step height to be $V = 0.9J$.
Comparing with Fig.~\ref{figideal}(a), one can see that the momentum peaks are broader, reflecting the larger loss $\gamma$ and the smaller value of $\sigma$. 

In Fig.~\ref{figreal}(b), we plot the momentum distribution for $\phi_i = 18^\circ$; the reflected signal is visible and the transmitted signal is smaller compared to the normal incidence case of Fig.~\ref{figreal}(a). This trend is clearly visible in Fig.~\ref{figreal}(c), where we plot the transmission spectrum as a function of the angle of incidence for a fixed $V = 0.9J$, showing a clear decrease of the transmission rate as the angle is increased. Here, we are again estimating the reflection by integrating the whole signal at $k_x < 0$ of $x < 0$ region, which suffers from the broadened large incident peak centered at $k_x > 0$. Another difficulty of having a small sample size is the unwanted reflection at the edges of the sample. In particular, the reflection at the top and bottom edges of the system, which are aligned horizontally, can cause significant intervalley scatterings, which can result in the further deviation of the numerical simulation from the Klein tunneling theory.

Finally, in Fig.~\ref{figreal}(d), we plot the transmission as a function of the step height $V/\omega$ for a fixed incident angle of $\phi_i = 30^\circ$. In spite of a significant smoothening of the dip, the overall typical features of Klein tunneling are still well visible.

Summing up, the results of Fig.~\ref{figreal} confirm the actual observability of the main qualitative signatures of Klein tunneling in models of photonic graphene with realistic parameters. These results should provide a clear guideline to undertake the experiment with state-of-the-art samples.

\section{Conclusion}
\label{sec:conclu}
We have shown that the Klein tunneling and negative refraction effects can be observed in driven-dissipative photonic systems with experimentally realistic parameters. In particular, we have proposed an experimental scheme to extract the transmittivity in driven-dissipative photonic systems. Our result can be not only useful in confirming known condensed matter theories which are difficult to directly observe in electronic systems, but can also open the way to new phenomena that are unique in photonic systems.

Firstly, the structure of the spinor wave function is clearly visible through the interference effect in the momentum space emission of photons. By choosing the relative phase of the sublattices via the wave vector of the external pump field, one is able to better isolate the physics of interest and possibly explore new subtle features of sublattice-dependent physics.

Secondly, an even more exciting direction of future research will consist of including optical nonlinearities characteristic of the polariton microcavities modeled in the present work. This would allow taking Klein tunneling, and also other effects characteristic of the spinor nature of the wave function, to the nonlinear regime.

\begin{acknowledgements}
We thank T. Jacqmin and M. Mili\'cevi\'c for stimulating discussions.
This work was supported by the ERC grants QGBE and Honeypol, by the EU-FET Proactive grant AQuS, Project No. 640800, by the Provincia Autonoma di Trento partially through the project ``On silicon chip quantum optics for quantum computing and secure communications -- SiQuro'', the Labex NanoSaclay project ICQOQS (Grant No. ANR-10-LABX-0035), and the French National Research Agency (ANR) project ”Quantum Fluids of Light” (ANR-16-CE30-0021).
\end{acknowledgements}

\end{document}